\documentclass[preprint,review,12pt]{revtex4}
\usepackage{graphicx}
\usepackage{dcolumn}
\usepackage{bm}
\usepackage[utf8]{inputenc}
\usepackage[T1]{fontenc}
\usepackage{mathptmx}
\usepackage{etoolbox}
\usepackage{color}
\usepackage{amsmath,amssymb}
\usepackage{braket}
\usepackage{ulem}

\makeatletter
\def\@email#1#2{%
 \endgroup
 \patchcmd{\titleblock@produce}
  {\frontmatter@RRAPformat}
  {\frontmatter@RRAPformat{\produce@RRAP{*#1\href{mailto:#2}{#2}}}\frontmatter@RRAPformat}
  {}{}
}%
\makeatother

\usepackage{multirow}
\usepackage{array}
\newcolumntype{C}[1]{>{\hfil}m{#1}<{\hfil}}
\makeatletter
\renewcommand{\theequation}{
    \thesection.\arabic{equation}}
    \@addtoreset{equation}{section}
\makeatother

\begin{document}

%\documentclass[preprint,review,12pt]{revtex4}
%\usepackage{graphicx}
%\usepackage{dcolumn}
%\usepackage{bm}
%\usepackage[utf8]{inputenc}
%\usepackage[T1]{fontenc}
%\usepackage{mathptmx}
%\usepackage{etoolbox}
%\usepackage{color}
%\usepackage{amsmath,amssymb}
%\usepackage{braket}
%\usepackage{ulem}
%\newcommand{\average}[1]{\ensuremath{\langle#1\rangle} }
%\makeatletter
%\def\@email#1#2{%
% \endgroup
% \patchcmd{\titleblock@produce}
%  {\frontmatter@RRAPformat}
%  {\frontmatter@RRAPformat{\produce@RRAP{*#1\href{mailto:#2}{#2}}}\%frontmatter@RRAPformat}
%  {}{}
%}%
%\makeatother
%\usepackage{pdfpages}
%\begin{document}

%\preprint{AIP/123-QED}

\title{Construction of Machine-Learning Interatomic Potential Under Heat Flux Regularization and Its Application to Power Spectrum Analysis for Silver Chalcogenides}

\author{Kohei Shimamura}
\email{shimamura@kumamoto-u.ac.jp.}
\thanks{Author to whom correspondence should be addressed}
\affiliation{Department of Physics, Kumamoto University, Kumamoto 860-8555, Japan}

\author{Akihide Koura}
\affiliation{Department of Physics, Kumamoto University, Kumamoto 860-8555, Japan}

\author{Fuyuki Shimojo}
\affiliation{Department of Physics, Kumamoto University, Kumamoto 860-8555, Japan}

%\date{\today}

\begin{abstract}
We propose a data-driven approach for constructing machine-learning interatomic potentials (MLIPs) trained under a regularization with the aim of avoiding nonphysical heat flux.
Specifically, we introduce a regularization term for the heat flux into the cost function of MLIPs to be minimized.
Since the treatment of heat flux using MLIPs with regularization can be decomposed into elemental contributions or conducted in frequency space, this approach is expected to be useful for investigating the origin of thermal conductivity obtained from the Green-Kubo formula.
However, the strength of regularization needs to be appropriately set because it may reduce not only the nonphysical part but also the intrinsic heat flux one.
To this end, we investigated the conditions for constructing MLIPs that can reproduce the power spectra of heat flux associated with the empirical interatomic potential of Ag$_2$Se, which consists of pairwise functions and do not contain a nonphysical heat flux.
The appropriate strength could be estimated from the variation of the magnitude of regularization term as well as root mean square errors for total potential energy, atomic force, and virial stress with respect to the strengths, without reference spectrum data.
As an application example, we explored the differences in power spectra between superionic and nonsuperionic conducting phases based on the heat flux regularization to MLIPs trained with the first-principles calculation data of Ag$_2$S.
Furthermore, our results demonstrate that training with the regularization improves the robustness of MLIPs as well as the reduction of the nonphysical heat flux.
\end{abstract}

\maketitle

\section{\label{sec:intro}Introduction}
The Green-Kubo (GK) method is an efficient tool for calculating thermal conductivity (TC)~\cite{Volz_2000,Kubo_1957,Green_1952} and can be applied to irregular systems, where the heat flux ${\bf{J}}_Q$ sampled using a molecular dynamics (MD) simulation is used:
\begin{eqnarray}
\kappa &=& \frac{1}{3k_{\rm B}T^2\varOmega}\int^{\infty}_{0} \langle{\bf{J}}_Q(t) \cdot {\bf{J}}_Q(0)\rangle dt
\label{eq:kappa},
\end{eqnarray}
where $k_{\rm B}$, $T$, and $\varOmega$ are the Boltzmann constant, temperature, and volume of the supercell, respectively.
The GK method has recently been combined with machine-learning interatomic potentials~\cite{Morawietz_2021} (MLIPs) to estimate the TCs of various materials.~\cite{Tisi_2021,Fan_2021,Verdi_2021,Li_2020,Huang_2019,Bosoni_2019}
MLIPs, which are trained with data obtained from first-principles molecular dynamics (FPMD) simulations, successfully replicate the high accuracy of FPMD while maintaining the low computational cost associated with conventional empirical interatomic potentials (EIPs).

However, a critical problem accompanies the analysis of the origin of TC using the GK method with MLIPs.
The MLIP is a many-body potential; thus, the total potential energy $E^{\rm MLIP}$ is defined as the sum of the individual atomic potential energies $\varepsilon^{\rm MLIP}_i$, 
\begin{eqnarray}
E^{\rm MLIP} = \sum^{N_{\rm atom}}_{i} \epsilon^{\rm MLIP}_i
\label{eq:energy},
\end{eqnarray}
where $N_{\rm atom}$ denotes the number of atoms in the system.
The set \{$\varepsilon^{\rm MLIP}_i$\} is indirectly determined through the training of $E^{\rm MLIP}$.
There are infinite sets of \{$\varepsilon^{\rm MLIP}_i$\} that generate the same value of $E^{\rm MLIP}$, i.e., $\varepsilon^{\rm MLIP}_i$ has a gauge (called an {\it{atomic gauge}}).~\cite{Kim_2021}
Thus, trained MLIPs with different initial weight parameters are likely to give the same $E^{\rm MLIP}$, but the set \{$\varepsilon^{\rm MLIP}_i$\} can be different for each MLIP.
According to Kim {\it et al.},~\cite{Kim_2021} the value of ${\bf{J}}_Q$ obtained from the interatomic potentials defined by Eq.~(\ref{eq:energy}) can be divided into ${\bf{J}}^{\rm net}_Q$ and ${\bf{J}}^{\rm nPhys}_Q$ due to the $atomic$ $gauge$:
\begin{eqnarray}
{\bf{J}}_Q = {\bf{J}}^{\rm net}_Q + {\bf{J}}^{\rm nPhys}_Q
\label{eq:JP}.
\end{eqnarray}
Only the net heat flux ${\bf{J}}^{\rm net}_Q$ contributes to the TC.
In contrast, it has been revealed that ${\bf{J}}^{\rm nPhys}_Q$ has no effect on the calculation of TC via Eq.~(\ref{eq:kappa}) because it is the total derivative with respect to time.~\cite{Kim_2021,Ercole_2016}
Accordingly, ${\bf{J}}^{\rm nPhys}_Q$ is a nonphysical heat flux.
The analysis based on the decomposition of ${\bf{J}}_Q$ becomes meaningless because of the presence of ${\bf{J}}^{\rm nPhys}_Q$, whose values would be different for each MLIP.
On the other hand, the EIPs that consist of only pairwise functions can define the atomic potential energy unambiguously, and it is therefore considered that the influence of ${\bf{J}}^{\rm nPhys}_Q$ can be avoided.
Previous studies with such pairwise EIPs succeeded in detailed analyses of the TC by decomposing ${\bf{J}}_Q$ into elemental/atomic components~\cite{Fujii_2020,Ouyang_2014} or in frequency space.~\cite{Fujii_2018,Zhou_2015,Zhou_2015_2,Saaskilahti_2014}
The presence of ${\bf{J}}^{\rm nPhys}_Q$ prevents the application of these useful analytical techniques to MLIPs.

We propose a solution to this problem with a data-driven approach, which provides a way to prevent ${\bf{J}}^{\rm nPhys}_Q$ from playing a role.
Specifically, a regularization term associated with ${\bf{J}}_Q$ is introduced into the cost function $C$ to be minimized during the training of MLIPs:
\begin{eqnarray}
C = C_{\rm0} + p_J|{\bf{J}}_Q|^2
\label{eq:RJ},
\end{eqnarray}
where $C_{\rm0}$ denotes the cost function without the regularization.
$C_{\rm0}$ includes the loss functions for the total potential energy, atomic force, and virial stress in the training of MLIPs.~\cite{Irie_2021,Fan_2021,Marques_2019,Zhang_2018,Novikov_2021}
The regularization term brings ${\bf{J}}_Q$ closer to 0 during training, thus reducing ${\bf{J}}^{\rm nPhys}_Q$.
However, if the regularization is too strong, ${\bf{J}}^{\rm net}_Q$ is also impaired, substantially reducing the accuracy of the MLIP.
Therefore, it is vital to adjust the coefficient $p_J$.

To verify the effectiveness of this approach to heat flux regularization, we used an EIP of Ag$_2$Se~\cite{Rino_1988} consisting of pairwise functions.
The EIP of Ag$_2$Se is physically superior, being able to describe the phase transitions between the liquid, superionic ($\alpha$), and nonsuperionic ($\beta$) phases.~\cite{Shimojo_1991,Rino_1988}
First, taking advantage of the low computational cost of EIPs, the heat flux autocorrelation functions (HFACFs) $\langle{\bf{J}}_Q(t) \cdot {\bf{J}}_Q(0)\rangle$ for $\beta$- and $\alpha$-Ag$_2$Se and the corresponding power spectra $S(\omega)$ based on the Wiener-Khintchine theorem were calculated and used as reference data.
In addition to $S(\omega)$ being affected by ${\bf{J}}^{\rm nPhys}_Q$, which helps to verify the effect of regularization, $S(0)$ matches the TC, and $S(\omega)$ for $\omega > 0$ represents the frequency distribution of phonons.~\cite{Mahdizadeh_2014,Zhang_2013}
Second, we investigated the condition of the coefficient $p_J$ of the regularization term under which the MLIPs trained with the EIP-MD data of $\beta$- and $\alpha$-Ag$_2$Se reproduced the reference power spectra.
Next, as an application, the results of the previous step were used to construct MLIPs trained with FPMD data for the $\beta$ and $\alpha$ phases of Ag$_2$S. 
Finally, we analyzed the difference between the power spectra of the two phases.
Notwithstanding the difference in the dynamics of the Ag atoms, the TCs of the two phases have been experimentally estimated as similar values of approximately 0.5 Wm$^{\mathchar`-1}$K$^{\mathchar`-1}$.~\cite{Hirata_2020,Matsunaga_2021}
An analysis of power spectra in frequency space would provide information on the characteristic phonon distributions in the two phases.
The results of the above-mentioned procedures might facilitate the development of silver chalcogenides such as Ag$_2$Se and Ag$_2$S as thermoelectric materials and thermoelectric diode devices.

In this study, we adopted the artificial neural network (ANN) potential~\cite{Artrith_2016,Irie_2021} as the MLIP.
We recently proposed a computational framework for calculating TC by the GK method using ANN potentials, reporting the possibility of estimating the TCs of $\beta$- and $\alpha$-Ag$_2$Se with high accuracy.~\cite{Takeshita_2022,Shimamura_2021,Shimamura_2020}
In the present study, we further improved the framework by the heat flux regularization.

In section~\ref{sec:methods}, we explain our computational methodologies, such as the construction of the ANN potential, the formulas for the heat flux and the power spectrum, the creation of training data for Ag$_2$Se and Ag$_2$S, the cost function, the regularization term, and the calculation of TC and the power spectrum.
In section~\ref{sec:results}, we first discuss the estimation of the appropriate coefficient $p_J$ of the regularization term through the reproduction of power spectra from the EIP of Ag$_2$Se.
Second, we apply this approach to Ag$_2$S without reference data and mention the difference between the power spectra of the $\beta$ and $\alpha$ phases.
We finally present our conclusions in section~\ref{sec:conclusion}.

\section{\label{sec:methods} Computational Methods}
\subsection{ANN Potential}
\label{SS:ann}
The ANN potentials considered in this study were constructed using the Aenet code, which is a training code developed by Artrith {\it et al.}~\cite{Artrith_2016}
The ANN potential constructed by the Aenet code comprised feed forward neural networks (FFNNs) created for each atomic species, comprising two hidden layers with ten nodes.
The scaled hyperbolic tangent~\cite{Artrith_2016} was used as the activation function of hidden layers.
The total potential energy $E^{\rm ANN}$ is defined as the sum of the atomic potential energies \{$\epsilon^{\rm ANN}_i$\} output from FFNNs for all atoms in the system,~\cite{Behler_2007} i.e., via Eq.~(\ref{eq:energy}).

A numerical descriptor {\bf{$\sigma_i$}} that characterizes the atomic structure around each atom was used as the input for the ANN potential to obtain $\epsilon^{\rm ANN}_i$.
Among the various methods for creating such descriptors, Artrith {\it et al.} proposed the Chebyshev descriptor that exploits the recursivity of the Chebyshev polynomial.~\cite{Artrith_2017}
The seventeen radial and five angular Chebyshev descriptors were used with the cutoff radii of 8 {\AA} and 5 {\AA}, respectively.

\subsection{\label{ssec:GKmethod}Green-Kubo method}
\subsubsection{\label{sssec:hf} Heat flux formula}
The heat flux ${\bf{J}}_Q$ in Eq.~(\ref{eq:kappa}) for the EIP of Ag$_2$Se and the ANN potential is defined as~\cite{Shimamura_2020}
\begin{eqnarray}
{\bf{J}}_Q &=& \sum^{N_{\rm atom}}_{i} (t_i + \epsilon_i){\bf{v}}_i \
                + \sum^{N_{\rm atom}}_{i} {\rm W}_i{\bf{v}}_i \nonumber \\
              &-& \sum^{2}_{\mu} h_{\mu} \sum^{N_{\mu}}_{i \in \mu} {\bf{v}}_i \
\label{eq:hflux},
\end{eqnarray}
where $t_i$, $\epsilon_i$, ${\bf{v}}_i$, and ${\rm W}_i$ are the kinetic energy, potential energy, velocity, and virial tensor for the $i$th atom, respectively.
In addition, $h_{\mu}$ and $N_{\mu}$ denote the partial enthalpy and the number of atoms for the $\mu$th component in the system ($\mu \in$ \{Ag, Se\} for Ag$_2$Se and $\mu \in$ \{Ag, S\} for Ag$_2$S), respectively.
The formulas for $\epsilon_i$ and ${\rm W}_i$ for the EIP of Ag$_2$Se are shown in Section I of the Supplementary Materials (SM).
${\rm W}_i$ for the ANN potential with the Chebyshev descriptor is defined as~\cite{Shimamura_2021}
\begin{eqnarray}
{\rm W}^{\rm ANN}_i = \sum_{j \ne i} {\bf{r}}_{ij} \otimes \frac{\partial \epsilon^{\rm ANN}_j}{\partial {\bf{r}}_{ji}}
\label{eq:virial},
\end{eqnarray}
where ${\bf{r}}_{ij}$ $\equiv$ ${\bf{r}}_{j} - {\bf{r}}_{i}$ and
${\bf{r}}_{i}$ denotes $i$th atomic coordinate.
Note that the summation of ${\rm W}_i$ over $N_{\rm atom}$ coincides with the total virial tensor ${\rm W}$ defined by the virial theorem.
The potential part of the virial stress tensor ${\rm \Pi_{\rm pot}}$ is calculated by dividing ${\rm W}$ by the volume of the supercell $\varOmega$.
The third term on the right-hand side in Eq.~(\ref{eq:hflux}) is introduced to correct the difference in the particle number flux between each component,~\cite{Perronace_2002} which is necessary in binary or other multiple system.

In the GK method, decomposing ${\bf{J}}_Q$ into constituent parts enables us to estimate their respective contributions to the TC.
To elucidate the elemental contributions from components 1 and 2 denoted as $\mu_1$ and $\mu_2$, respectively, ${\bf{J}}_Q$ in Eq.~(\ref{eq:hflux}) is decomposed into ${\bf{J}}_{\mu1}$ and ${\bf{J}}_{\mu2}$.
Incidentally, ${\bf{J}}_Q$ can be also divided into the kinetic (convective) and potential (conductive) parts denoted as ${\bf{J}}_{\rm K}$ and ${\bf{J}}_{\rm P}$, respectively.
${\bf{J}}_{\rm K}$ and ${\bf{J}}_{\rm P}$ consist of terms in Eq.~(\ref{eq:hflux}) containing $t_i$ and $\epsilon_i$ and containing ${\rm W}_i$, respectively.

\subsubsection{\label{sssec:ps} Power spectrum of heat flux}
According to the Wiener-Khintchine theorem, the Fourier transformation of the time series of the HFACF produces the power spectrum $S(\omega)$:
\begin{eqnarray}
S(\omega) &=& \int^{\infty}_{-\infty} e^{i\omega t} \langle{\bf{J}}_Q(t) \cdot {\bf{J}}_Q(0)\rangle dt \nonumber \\
&=& 2\int^{\infty}_{0} \cos(\omega t) \langle{\bf{J}}_Q(t) \cdot {\bf{J}}_Q(0)\rangle dt
\label{eq:ps}.
\end{eqnarray}
Furthermore, the elemental and kinetic/potential contributions  of $S(\omega)$ can also be calculated using the decomposed heat fluxes mentioned in section~\ref{sssec:hf}: $S(\omega) = S_{\mu1\mathchar`-\mu1}(\omega) + S_{\mu1\mathchar`-\mu2}(\omega) + S_{\mu2\mathchar`-\mu1}(\omega) + S_{\mu2\mathchar`-\mu2}(\omega)$ and $S(\omega) = S_{\rm K\mathchar`-K}(\omega) + S_{\rm K\mathchar`-P}(\omega) + S_{\rm P\mathchar`-K}(\omega) + S_{\rm P\mathchar`-P}(\omega)$.

\subsection{\label{ssec:eipdata}Generation of Ag$_2$Se training data by MD simulation with EIP}
The MD simulations in this paper were executed using the QXMD~\cite{Shimojo_2019} code.
In this section, we use the EIP proposed by Rino {\it{et al}}.,~\cite{Rino_1988} comprising three pairwise potential functions: repulsive, dielectric dipole interaction, and Coulomb interaction terms (see Section I of the SM).

\subsubsection{\label{sssec:bag2se}$\beta$-Ag$_2$Se at 300 K}
We generated the training data of $\beta$-Ag$_2$Se at 300 K by performing MD simulations with the EIP of Ag$_2$Se.~\cite{Rino_1988}
The unit cell of $\beta$-Ag$_2$Se reproduced by the EIP has an orthorhombic structure consisting of four Ag$_2$Se molecules with the following lattice parameters: $a = 4.29$ \AA, $b = 6.82$ \AA, and $c = 8.25$ \AA.~\cite{Shimojo_1991}
Using the $8\times2\times2$ unit cell structure consisting of 256 Ag and 128 Se atoms under periodic boundary conditions, an MD simulation with the canonical ensemble~\cite{MP_52_255} (hereinafter called $NVT$-MD) was performed with 100,000 steps at 300 K.
The time step $\Delta t$ was set to 2.42 fs. 
A total of 1,000 MD steps were used as training data by extracting every 100 steps from the 100,000 MD steps to remove similar structures.
The atomic coordinates \{${\bf{r}}^{\beta\rm EIP}_{I,i}$\}, atomic velocities \{${\bf{v}}^{\beta\rm EIP}_{I,i}$\}, total potential energy $E^{\beta\rm EIP}_I$, atomic force \{${\bf{F}}^{\beta\rm EIP}_{I,i}$\}, virial tensor ${\rm W}^{\beta\rm EIP}_{I}$, and supercell volume $\varOmega^{\beta\rm EIP}_I$, were contained in the $I$th MD step data.

\subsubsection{\label{sssec:aag2se}$\alpha$-Ag$_2$Se at 500 K}
The data generated using the EIP of Ag$_2$Se in our previous study~\cite{Shimamura_2020} were used in this study.
Using the experimentally obtained number density for $\alpha$-Ag$_2$Se at 500 K (0.0461 \AA$^{\mathchar`-3}$),~\cite{Okazaki_1967} we prepared a system with 256 Ag and 128 Se atoms in a cubic supercell under periodic boundary conditions.
The length of one side of the supercell was 20.271~\AA.
For the configuration, an $NVT$-MD simulation using the EIP of Ag$_2$Se was performed with 100,000 steps at 500 K.
The time step $\Delta t$ was set to 2.42 fs. 
A total of 1,000 MD steps were used as training data by extracting every 100 steps.
The atomic coordinates \{${\bf{r}}^{\alpha\rm EIP}_{I,i}$\}, atomic velocities \{${\bf{v}}^{\alpha\rm EIP}_{I,i}$\}, total potential energy $E^{\alpha\rm EIP}_I$, atomic force \{${\bf{F}}^{\alpha\rm EIP}_{I,i}$\}, virial tensor ${\rm W}^{\alpha\rm EIP}_{I}$, and supercell volume $\varOmega^{\alpha\rm EIP}_I$, were contained in the $I$th MD step data.

\subsection{\label{ssec:fpdata}Generation of training data of Ag$_2$S by FPMD simulation}

\subsubsection{\label{sssec:bag2s}$\beta$-Ag$_2$S at 300 K}
We created training data of $\beta$-Ag$_2$S at 300 K by performing FPMD simulations.
The electronic states were calculated by the projector augmented wave method~\cite{Blochl_1994,Kresse_1999} within the framework of density functional theory (DFT).~\cite{Hohenberg_1964,Kohn_1965}
Projector functions were generated for the 4$d$, 5$s$, and 5$p$ states of Ag and for the 3$s$, 3$p$, and 3$d$ states of S.
The Perdew-Burke-Ernzerhof generalized gradient approximation~\cite{Perdew_1996} was employed for the exchange correlation energy.
To correctly represent the electronic states in localized $d$ orbitals of Ag, the DFT+U method~\cite{Anisimov_1997} with the effective parameter for the Coulomb interaction $U_{\rm eff} = 6.0$ eV~\cite{Fukushima_2019_2, Santamaria_2012} was used.
The empirical correction of the van der Waals interaction by the DFT-D approach~\cite{Grimme_2006} was employed.
The plane wave cutoff energies were 20 and 200 Ry for the electronic pseudo-wave function and pseudo-charge density, respectively.
The energy functional was minimized iteratively using a preconditioned conjugate-gradient method~\cite{Shimojo_2001}.
The $\Gamma$ point was used for Brillouin zone sampling.

The unit cell of $\beta$-Ag$_2$S has a monoclinic structure consisting of four Ag$_2$S groups with lattice parameters: $a = 4.231$ \AA, $b = 6.930$ \AA, $c = 9.526$ \AA, and $\beta = 125.48^\circ$.~\cite{Sadanaga_1967}
Considering that the structure is a slightly distorted bcc lattice of S atoms, 32 unit cells were arranged to construct an atomic configuration close to cubic structure, containing 256 Ag and 128 S atoms.
The lattice parameters of supercell were $a = 19.548$ \AA, $b = 19.052$ \AA, $c = 19.548$ \AA, $\alpha = 90.871^\circ$, $\beta = 90.312^\circ$, and $\gamma = 90.871^\circ$.
Using the configuration under periodic boundary conditions, an FPMD simulation with the isothermal and isobaric ensemble (hereinafter called $NPT$-FPMD) was performed with 3,000 steps at 300 K and 0 GPa.
The time step $\Delta t$ was set to 1.21 fs. 
A total of 1,000 MD steps were used as training data by extracting every 3 steps.
The $I$th MD step data contained the atomic coordinates \{${\bf{r}}^{\beta\rm FP}_{I,i}$\}, atomic velocities \{${\bf{v}}^{\beta\rm FP}_{I,i}$\}, total potential energy $E^{\beta\rm FP}_I$, atomic force \{${\bf{F}}^{\beta\rm FP}_{I,i}$\}, virial tensor ${\rm W}^{\beta\rm FP}_{I}$, and supercell volume $\varOmega^{\beta\rm FP}_{I}$.

\subsubsection{\label{sssec:aag2s}$\alpha$-Ag$_2$S at 600 K}
We created training data of $\alpha$-Ag$_2$S at 600 K by performing FPMD simulations.
The electronic state calculation was performed with the same setting as that for $\beta$-Ag$_2$S described in the previous section~\ref{sssec:bag2s}.
Using the experimental lattice constant of  $\alpha$-Ag$_2$S,~\cite{Sadovnikov_2018} we prepared a system with 256 Ag and 128 S atoms in a cubic supercell under periodic boundary conditions, where the length of one side of the supercell was 19.496 \AA.

With that configuration, an $NPT$-FPMD simulation was performed with 3,000 steps at 600 K and 0 GPa.
The time step $\Delta t$ was set to 1.21 fs. 
A total of 1,000 MD steps were used as training data by extracting every 3 steps.
The $I$th MD step data contained the atomic coordinates \{${\bf{r}}^{\alpha\rm FP}_{I,i}$\}, atomic velocities \{${\bf{v}}^{\alpha\rm FP}_{I,i}$\}, total potential energy $E^{\alpha\rm FP}_I$, atomic force \{${\bf{F}}^{\alpha\rm FP}_{I,i}$\}, virial tensor ${\rm W}^{\alpha\rm FP}_{I}$, and supercell volume $\varOmega^{\alpha\rm FP}_{I}$.

\subsection{\label{ssec:training}Training of ANN potential}
\subsubsection{\label{sssec:cost}Cost function}
We define the following cost function $C$ for training ANN potentials. 
It comprises three loss functions associated with the total potential energy (first term), atomic force (second term), and virial (third term), and a regularization term associated with the heat flux (fourth term):
\begin{eqnarray}
C &=& \frac{p_E}{2}\frac{1}{N_I}\sum^{N_I}_{I} \left(  \frac{   E^{\rm ANN}_I - E^{\rm Ref}_I  }{N_{{\rm atom},I}} \right)^2 \nonumber \\
&+& \frac{p_F}{2}\frac{1}{N_I}\sum^{N_I}_{I} \frac{1}{3N_{{\rm atom},I}} \sum^{N_{{\rm atom},I}}_{i}\ 
    \left({\bf{F}}^{\rm ANN}_{I,i} - {\bf{F}}^{\rm Ref}_{I,i} \right)^2 \nonumber \\
&+& \frac{p_W}{2}\frac{1}{N_I}\sum^{N_I}_{I} \frac{1}{6} \sum^{6}_{j}\ 
    \left(  \frac{{W}^{\rm ANN}_{I,j} - {W}^{\rm Ref}_{I,j}}{N_{{\rm atom},I}} \right)^2 \nonumber \\
&+& \frac{p_J}{2}\frac{1}{N_I}\sum^{N_I}_{I}
    \frac{1}{3N_{{\rm atom},I}}
    \left(  {\bf{J}'}_{{Q},I} \right)^2
\label{eq:cost},
\end{eqnarray}
where $N_I$ denotes the number of training data and ${\bf{F}}^{\rm ANN}_{i}$ is given by~\cite{Shimamura_2021} 
\begin{eqnarray}
{\bf{F}}^{\rm ANN}_{i} 
&=& \sum^{N_{\rm atom}}_j {\bf{F}}^{\rm ANN}_{ij} \nonumber \\
&=& -\sum^{N_{\rm atom}}_j \left[\frac{\partial \epsilon^{\rm ANN}_i}{\partial {\bf{r}}_{ji}} + \frac{\partial \epsilon^{\rm ANN}_j}{\partial {\bf{r}}_{ji}} \right]
\label{eq:pforce}.
\end{eqnarray}
The pairwise force  
${\bf{F}}^{\rm ANN}_{ij}$
satisfies Newton's third law.~\cite{Fan_2015}
The prime heat flux ${\bf{J}'}_{{Q}}$ in the fourth term of Eq.~(\ref{eq:cost}) will be described in the next section~\ref{sssec:regu}.
Because these terms differ in dimension and size, $p_E$, $p_F$, $p_W$, and $p_J$ are introduced as adjustable parameters.
The symbols with ``Ref'' denote training data, i.e., those labeled ``$\beta$EIP'', ``$\alpha$EIP'', ``$\beta$FP'', or ``$\alpha$FP'' and explained in sections~\ref{ssec:eipdata} and ~\ref{ssec:fpdata}.

With each "Ref" data, we performed training on 80\% of them and the remaining 20\% were used for testing.
Hereafter, the former and latter are called the Train and Test data sets, respectively.
The Levenberg-Marquardt method~\cite{Artrith_2016} was adopted for the optimization process.
Further details of the training procedures undertaken to minimize the cost function are provided in Section II of the SM, except for $p_J$, whose values are presented in section~\ref{sssec:regu}.

\subsubsection{\label{sssec:regu} Regularization term of heat flux}
The definition of ${\bf{J}'}_{{Q}}$ in Eq.~(\ref{eq:cost}) is different from that of ${\bf{J}}_{{Q}}$ in Eq.~(\ref{eq:hflux}):
\begin{eqnarray}
{\bf{J}'}_Q &=& 
\sum^{N_{\rm atom}}_{i} {\rm W}^{\rm ANN}_i{\bf{u}}^{\rm Ref}_i
\label{eq:rehflux},
\end{eqnarray}
where ${\bf{u}}^{\rm Ref}_i$ is a renormalized velocity of the $i$th atom calculated from the atomic velocities \{${\bf{v}}^{\rm Ref}_{i}$\} in the training data.
If the $i$th atom belongs to the $\mu$th component, ${\bf{u}}^{\rm Ref}_i$ is calculated by subtracting the average velocity of all atoms belonging to the $\mu$th component from ${\bf{v}}^{\rm Ref}_{i}$:
\begin{eqnarray}
{\bf{u}}^{\rm Ref}_i &=& 
{\bf{v}}^{\rm Ref}_i
-\frac{1}{N_{\mu}}\sum^{N_{\mu}}_{j \in \mu} {\bf{v}}^{\rm Ref}_j
\label{eq:revel}.
\end{eqnarray}
The reason for using the renormalized velocity is to play an alternative role to the third term of Eq.~(\ref{eq:hflux}).
Marcolongo {\it et al.} recently showed that replacing atomic velocities in the heat flux formula with the renormalized velocities approximately corresponds to the introduction of the correction embodied in the third term.~\cite{Marcolongo_2020}
Since the calculation of enthalpy requires a huge number of statistics and obtaining it from a small amount of training data causes an error in the calculation of heat flux, their approach was employed in the regularization term.
The reason for considering only the potential part of the heat flux in Eq.~(\ref{eq:rehflux}) was that its contribution to the TC was dominant compared to that of the kinetic part in our previous studies regarding silver chalcogenides.~\cite{Takeshita_2022,Shimamura_2021}

The effect of the regularization term on training is controlled by the magnitude of its coefficient $p_J$.
In this study, in addition to $p_j = 0.0$, we mainly set $p_J = 10^{\mathchar`-4}$, $10^{\mathchar`-3}$, $10^{\mathchar`-2}$, or $10^{\mathchar`-1}$ to construct ANN potentials and compare their results.

Furthermore, we defined the magnitude of the heat flux for regularization, $\Delta J_Q$, to evaluate the value of the heat flux in training as: $\Delta J_Q$ = $\sqrt{\frac{1}{N_I}\sum^{N_I}_{I}
    \frac{1}{3N_{{\rm atom},I}} \left(  {\bf{J}'}_{{Q},I} \right)^2}$ (eV$\cdot${\AA}/ps).

\subsection{\label{ssec:epMD}Calculation of power spectrum of heat flux}
In this study, the power spectra $S(\omega)$ for the following five potential models were computed:
\begin{enumerate}
  \setcounter{enumi}{-1}
  \item EIP of Ag$_2$Se for creating reference data of $\beta$- and $\alpha$-Ag$_2$Se. 
  \item ANN potentials trained with EIP-MD data of $\beta$-Ag$_2$Se. 
  \item ANN potentials trained with EIP-MD data of $\alpha$-Ag$_2$Se. 
  \item ANN potentials trained with FPMD data of $\beta$-Ag$_2$S.
  \item ANN potentials trained with FPMD data of $\alpha$-Ag$_2$S.
\end{enumerate}
Using the same initial atomic configuration, temperature, and time step as those for the training data, an $NVT$-MD simulation was performed. 
Subsequently, a 1,000,000 MD-step simulation with the microcanonical ensemble (hereinafter called $NVE$-MD) was performed for sampling ${\bf{J}}_Q$ as defined in Eq.~(\ref{eq:hflux}).
The power spectra $S(\omega)$ were finally computed using Eq.~(\ref{eq:ps}) with an upper limit of integration of 2 ps.

As described in section~\ref{sec:intro}, ${\bf{J}}_Q$ for the ANN potential is considered to depend on its initial weight parameters.
Therefore, for statistical evaluation, five ANN potentials with different initial weight parameters at each $p_J$ were constructed.
These potentials are denoted as ANN1-ANN5 hereafter.
The same initial values were used if the names of the potentials matched, even when the values of $p_J$ were different.

\section{\label{sec:results}Results and Discussion}

\subsection{\label{ssec:kappaemp}The reference power spectrum data obtained with EIP of Ag$_2$Se}
\begin{figure}[b]
\begin{center}
  \includegraphics[width=8.5cm]{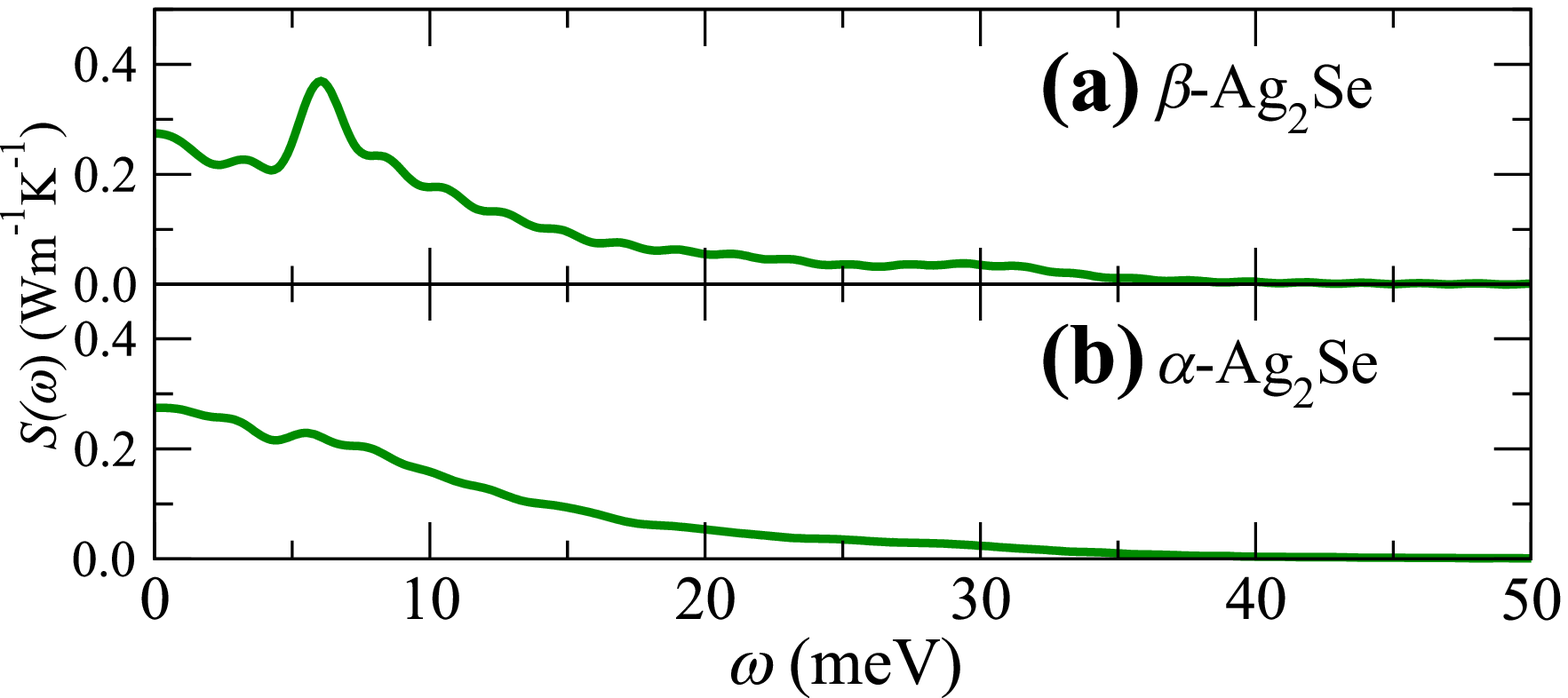}
  \caption{\label{Fig1}Power spectra of heat flux $S(\omega)$ obtained by the empirical interatomic potential (EIP) of Ag$_2$Se for (a) $\beta$- and (b) $\alpha$-Ag$_2$Se.}
\end{center}
\end{figure}

The power spectra $S(\omega)$ of $\beta$- and $\alpha$-Ag$_2$Se calculated using the EIP are displayed in Fig.~\ref{Fig1}.
Although the obtained TCs are almost the same at $\sim$0.3 Wm$^{\mathchar`-1}$K$^{\mathchar`-1}$, the profiles are distinctly different.
The power spectrum of the $\beta$ phase is characterized by a peak near 7 meV.
This peak is due to the $S_{\rm Ag\mathchar`-Ag}(\omega)$, as shown in Fig.~S1 of the SM.
The time variation of the corresponding HFACF and the cumulative TC $\kappa(t)$ defined in Eq.~(III.1) are shown in Fig.~S2 of the SM.

\subsection{\label{ssec:EIPresults} 
Results of ANN potentials trained with EIP-MD data}

\subsubsection{\label{sssec:re-bAg2Se}Power spectra of $\beta$-Ag$_2$Se and how to estimate the appropriate $p_J$} 
\begin{figure*}[b]
\begin{center}
  \includegraphics[width=15cm]{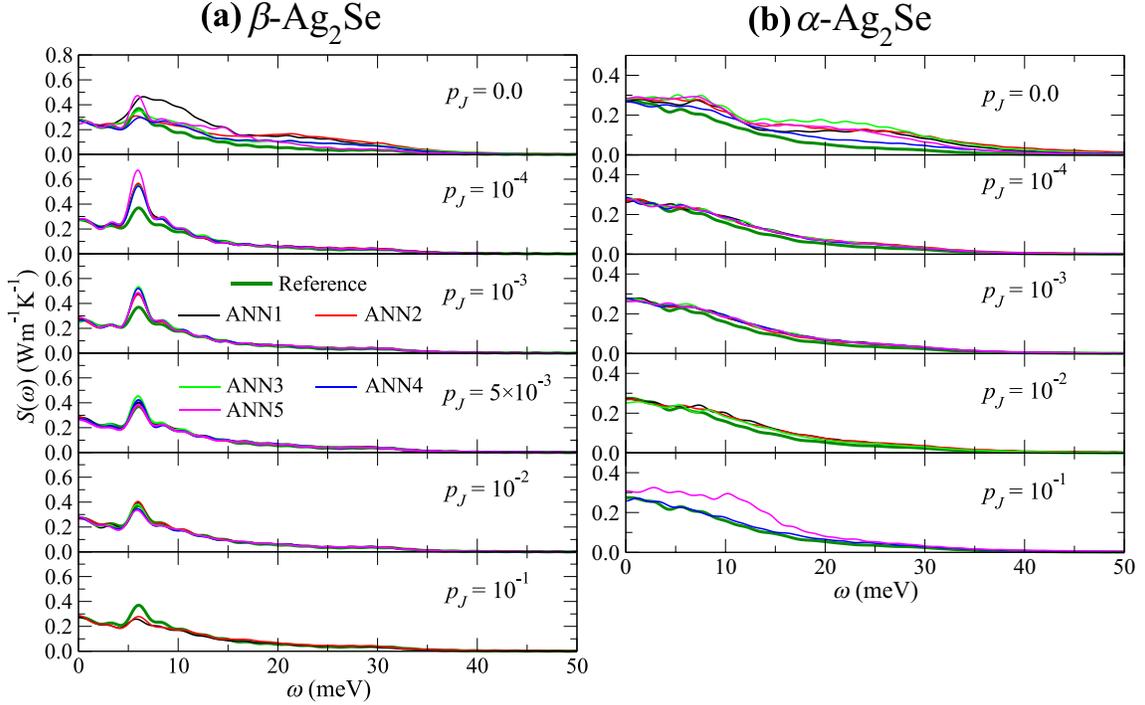}
  \caption{\label{Fig2}$S(\omega)$ obtained by the EIP of Ag$_2$Se (Reference, dark green) and trained ANN potentials with the heat flux regularization under each $p_J$ for (a) $\beta$- and (b) $\alpha$-Ag$_2$Se. ANN potentials (ANN1-ANN5) are trained with five different initial weight parameters (identified by colors in the figures above).}
\end{center}
\end{figure*}

The dependence of the $\beta$-Ag$_2$Se power spectra on the regularization coefficient $p_J$ is shown in Fig.~\ref{Fig2}(a), where the results of five ANN potentials (ANN1-ANN5) with different initial weight parameters are included.
The corresponding HFACFs and $\kappa(t)$ are shown in Figs.~S3(a) and~S3(c) of the SM.
With $p_J$ = 0.0 (i.e., without the regularization), the values of the TCs are almost the same (at $\sim$0.3 Wm$^{\mathchar`-1}$K$^{\mathchar`-1}$) for the five ANN potentials, whereas the spectra differ.
The profiles from 0 to 5 meV indicate that the power spectrum $S$($\omega$) of each of the ANN potentials lies close to the reference spectrum in this range; however, above 10 meV the match is poor.
As mentioned in section~\ref{sec:intro}, this occurs because ${\bf{J}}^{\rm nPhys}_Q$ is different for each ANN potential.
In contrast, with $p_J = 10^{\mathchar`-4}$, $10^{\mathchar`-3}$, and $10^{\mathchar`-2}$, the power spectra show very similar profiles among all the ANN potentials, although the peak heights at 7 meV for 10$^{\mathchar`-4}$ and 10$^{\mathchar`-3}$ were not identical.
The results with $p_J$ = 10$^{\mathchar`-2}$ were in good agreement with the reference spectrum for all five ANN potentials.
This result implies that ${\bf{J}}^{\rm nPhys}_Q$ was successfully reduced by the heat flux regularization.

Training with a much stronger regularization (i.e., $p_J$ = $10^{\mathchar`-1}$) not only renders the power spectra incompatible with the reference spectrum (evidenced by the lower peak heights at 7 meV), but also makes it difficult to perform long-term MD simulations.
Three of the five ANN potentials were unable to complete the 1,000,000-step $NVE$-MD simulations.
Hence, the corresponding spectra derived from ANN3-ANN5 are not depicted in Fig.~\ref{Fig2}(a).
The root mean square errors (RMSEs) for the total potential energy ($\Delta E$), atomic force ($\Delta F$), and virial stress from the contribution of the potential energy ($\Delta \varPi_{\rm pot}$), averaged over the five ANN potentials, are illustrated in Figs.~\ref{Fig3}(a)-\ref{Fig3}(c), as a function of $p_J$.
The specific formulas for the RMSEs are provided in Section II C of the SM.
The RMSEs for the Train and Test data are almost identical, suggesting that no overfitting occurs.
In addition, the magnitude of the heat flux for regularization $\Delta J_{Q}$ obtained from the training is shown as a function of $p_J$ for $\beta$-Ag$_2$Se in Fig.~\ref{Fig3}(d).
The averages and error bars (standard deviations) of the RMSEs for $p_J$ = $10^{\mathchar`-1}$ have much larger values compared to those for the other values of $p_J$, indicating that the accuracies of the ANN potentials worsened drastically as $p_J$ approached $10^{\mathchar`-1}$.
Conversely, $\Delta J_{Q}$ was the smallest when $p_J$ =  $10^{\mathchar`-1}$.
We hypothesize that the effect of strong regularization on ${\bf{J}}^{\rm net}_Q$, the net heat flux part, reduced the robustness of the ANN potentials.

These results provide a foundation for finding the appropriate $p_J$ without resorting to reference data.
$\Delta J_{Q}$ decreases monotonically as $p_J$ increases.
This behavior was the opposite of trend displayed by the RMSEs.
The optimal value of $p_J$ would achieve moderately low values for both $\Delta J_{Q}$ and  the RMSEs.
Note that stronger regularization results in a greater difference between the values of $\Delta J_{Q}$ obtained for the Train and Test data.
This result may also be used to estimate the appropriate value of $p_J$ because the difference between the values of $\Delta J_{Q}$ for the Train and Test data is expected to increase when excessive regularization is imposed.

We sought an optimal value of $p_J$ for $\beta$-Ag$_2$Se as follows.
As $\Delta E$ for the regularization with $p_J$ = 10$^{\mathchar`-2}$ is similar to that for $p_J$ = 10$^{\mathchar`-1}$ but relatively large compared to that of 10$^{\mathchar`-3}$, the optimum value for $p_J$ is inferred to lie below 10$^{\mathchar`-2}$.
Therefore, we additionally trained ANN potentials with $p_J$ = 5$\times$10$^{\mathchar`-3}$.
While $\Delta J_{Q}$ shows very little change for $p_J$ = 5$\times$10$^{\mathchar`-3}$ and 10$^{\mathchar`-2}$, the values of $\Delta E$, $\Delta F$, and $\Delta \varPi_{\rm pot}$ are all smaller for $p_J$ = 5$\times$10$^{\mathchar`-3}$ than for 10$^{\mathchar`-2}$; they also display power spectra close to the reference spectrum.
In this way, we could determine an optimal value of $p_J$ by assessing the RMSEs and $\Delta J_{Q}$ as functions of $p_J$, even if there are no reference data.
%We therefore propose that an optimal value of $p_J$ may be determined by assessing the RMSEs and $\Delta J_{Q}$ as functions of $p_J$, even if there are no reference data.
Although the RMSEs with $p_J$ = 5$\times$10$^{\mathchar`-3}$ were not minimal, we confirmed that the ANN potentials almost completely reproduced the atomic structure and ion dynamics, such as the radial distribution functions $g(r)$ and the mean squared displacements (MSDs) of the EIP of Ag$_2$Se, as shown in Figs.~S4(a) and~S5(a) of the SM.

\subsubsection{\label{sssec:re-aAg2Se}Power spectra of $\alpha$-Ag$_2$Se}
\begin{figure}[b]
\begin{center}
  \includegraphics[width=8.5cm]{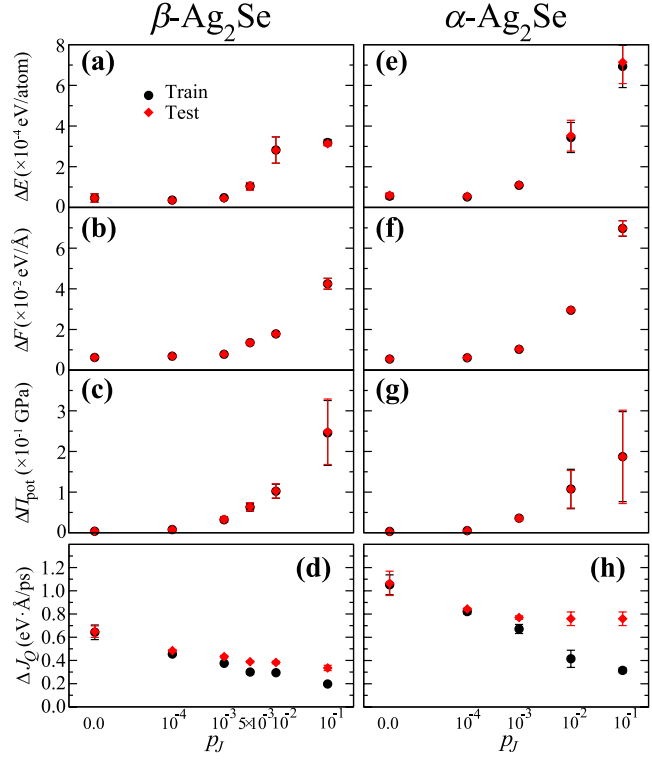}
  \caption{\label{Fig3} Averaged root mean square errors (RMSEs) and corresponding standard deviations (error bars) of the total potential energy ($\Delta E$), atomic force ($\Delta F$), and virial stress from the contribution of the potential energy ($\Delta \varPi_{\rm pot}$) over the five trained ANN potentials for (a-c) $\beta$- and (e-g) $\alpha$-Ag$_2$Se.
  The values of the magnitudes of heat flux for regularization ($\Delta J_Q$) for (d) $\beta$- and (h) $\alpha$-Ag$_2$Se are also shown. Black circles and red diamonds represent results obtained for the Train and Test data sets, respectively.}
\end{center}
\end{figure}

The power spectra of $\alpha$-Ag$_2$Se are shown in Fig.~\ref{Fig2}(b).
The corresponding HFACFs and $\kappa(t)$ are shown in Figs.~S3(b) and~S3(d) of the SM.
Qualitatively, the results are similar to those obtained for $\beta$-Ag$_2$Se as discussed in the previous section.
With $p_J$ = 0.0, the power spectra for different ANN potentials were not consistent, as expected.
All the spectra lay above the reference spectrum across the entire frequency range.
Conversely, the training with $p_J$ = 10$^{\mathchar`-4}$ or 10$^{\mathchar`-3}$ generated improved spectra, in good agreement with the reference spectrum.
With stronger regularization with $p_J$ = 10$^{\mathchar`-2}$ and 10$^{\mathchar`-1}$, $NVE$-MD simulations for 1,000,000 steps could be performed only with three (ANN1-ANN3) and two (ANN4 and ANN5) out of five ANN potentials, respectively.
With reference to the RMSEs displayed in Figs.~\ref{Fig3}(e)-\ref{Fig3}(g), the accuracy of the ANN potentials became considerably poor from $p_J$ = 10$^{\mathchar`-3}$ to 10$^{\mathchar`-2}$.
As shown in Fig.~\ref{Fig3}(h), the trend in $\Delta J_Q$ as a function of $p_J$ runs counter to that of the RMSEs. 
$\Delta J_Q$ for $p_J =$ 10$^{\mathchar`-2}$ shows a much larger deviation between the Train and Test data than for 10$^{\mathchar`-3}$, indicating that the strength of its regularization was excessive.  
The results discussed above imply that the optimal value for $\alpha$-Ag$_2$Se would be $p_J$ = 10$^{\mathchar`-3}$.
The $g$($r$) and MSDs obtained from the ANN potentials are comparable to those of the EIP of Ag$_2$Se as shown in Figs.~S4(b) and~S5(b).

\subsubsection{\label{sssec:val_heat}Improvement of robustness of ANN potential by heat flux regularization}
\begin{figure*}[b]
\begin{center}
  \includegraphics[width=15cm]{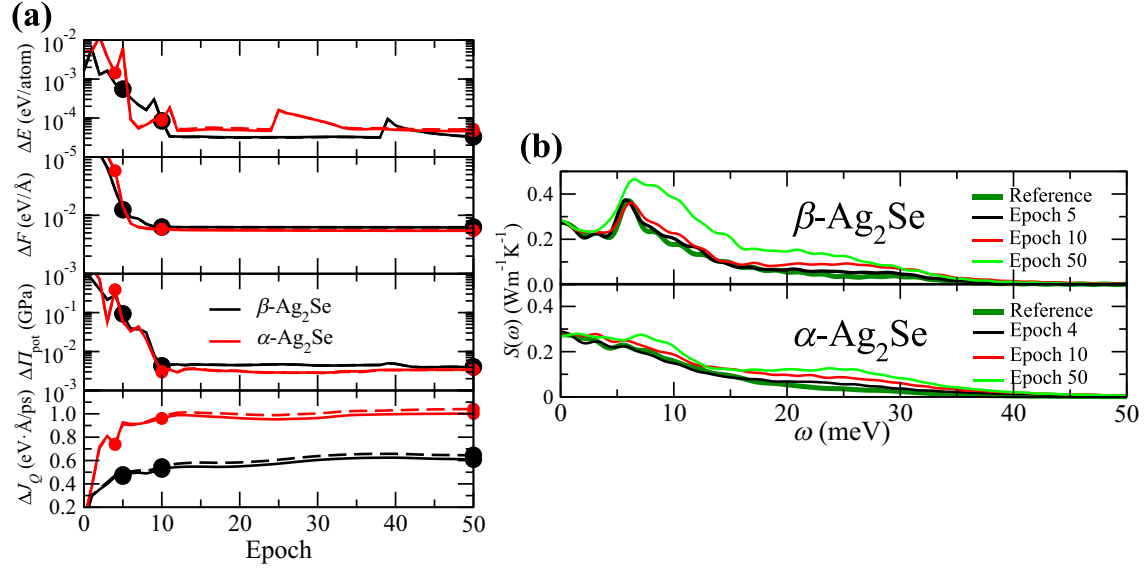}
  \caption{\label{Fig4}(a) The variation of $\Delta E$, $\Delta F$, $\Delta \varPi_{\rm pot}$, and $\Delta J_Q$ obtained during the training of an ANN potential for $\beta$- (black) and $\alpha$-Ag$_2$Se (red). The ANN potentials correspond to ANN1 for both phases in Fig.~\ref{Fig2}. The solid and dashed curves represent the RMSEs and $\Delta J_Q$ for the Train and Test data sets, respectively. (b) The corresponding $S(\omega)$ calculated using the ANN potentials at selected epochs marked with filled circles in (a), along with the results of the EIP of Ag$_2$Se (Reference). Epochs 4, 10, and 50 for $\beta$ and 5, 10, and 50 for $\alpha$ phases were selected.}
\end{center}
\end{figure*}

Tracking the variation in the heat flux value of $\Delta J_Q$ during training reveals interesting behavior.
The variations in $\Delta E$, $\Delta F$, $\Delta \varPi_{\rm pot}$, and $\Delta J_{Q}$ during training of one ANN potential (ANN1) without regularization for either $\beta$- or $\alpha$-Ag$_2$Se are illustrated in Fig.~\ref{Fig4}(a).
Up to 10 epochs, for both phases, the RMSEs rapidly decrease whereas, conversely, the  values of $\Delta J_Q$ increase.
After 10 epochs, the values of $\Delta E$ and $\Delta \varPi_{\rm pot}$ tended to be slightly smaller but remain approximately constant, whereas the values of $\Delta J_{Q}$ display an increasing trend.
The power spectra calculated using the ANN potentials at the selected three training epochs for each phase of Ag$_2$Se,
i.e., 5, 10, and 50 epochs for the $\beta$ phase and 4, 10, and 50 epochs for the $\alpha$ phase are presented in Fig.~\ref{Fig4}(b).
The power spectra for the $\beta$ and $\alpha$ phases at 5 and 4 epochs, respectively, were in good agreement with the reference spectra, but the deviation from the reference spectra increased with epoch number for both phases.
The corresponding TC is almost unchanged (see the power spectra at $S(0)$).
Therefore, it is considered that the small or nonexistent values of ${\bf{J}}^{\rm nPhys}_Q$ in the early epoch increased for subsequent epochs.
The increasing trend of $\Delta J_{Q}$ seen in Fig.~\ref{Fig4}(a) would reflect the increase in the value of ${\bf{J}}^{\rm nPhys}_Q$.
In addition, the RMSEs are higher at the epochs that give the agreement of the spectra with reference data.
The robustness of the ANN potentials after 10 epochs, when the RMSEs are almost at their lowest, is expected to deteriorate because their heat fluxes include ${\bf{J}}^{\rm nPhys}_Q$.

The increase in ${\bf{J}}^{\rm nPhys}_Q$ is ascribed to the fact that $E^{\rm ANN}$, ${\bf F}^{\rm ANN}_i$, and W$^{\rm ANN}$, that constitute the loss functions, are all sums of atomic components (i.e., $\varepsilon^{\rm ANN}_i$, ${\bf F}^{\rm ANN}_{ij}$, and W$^{\rm ANN}_{i}$, respectively) and that there are no training constraints on these components.
Because of the $atomic$ $gauge$ of $\varepsilon^{\rm ANN}_i$,~\cite{Kim_2021} there are infinite sets of \{$\varepsilon^{\rm ANN}_{i}$\}, \{${\bf F}^{\rm ANN}_{ij}$\}, and \{W$^{\rm ANN}_{i}$\} that produce the same values of $E^{\rm ANN}$, ${\bf F}^{\rm ANN}_i$, and W$^{\rm ANN}$, respectively.
During training, a set is sought for that will reduce the loss functions as much as possible, but the forcibly selected atomic components may have nonphysical values.
It can be considered that either overfitting or underfitting of the atomic components occurred.
%It has to be taken into consideration that either overfitting or underfitting of the atomic components occurred.
Furthermore, the heat flux is also composed of these atomic components (see Eqs.~(\ref{eq:hflux}) and~(\ref{eq:rehflux})).
${\bf F}^{\rm ANN}_{ij}$, which contains the same derivatives of $\varepsilon^{\rm ANN}_i$ as W$^{\rm ANN}_{i}$, would affect the heat flux indirectly through W$^{\rm ANN}_{i}$ (see Eqs.~(\ref{eq:pforce}) and~(\ref{eq:virial})).
Such a heat flux including nonphysical atomic components is considered to have a large ${\bf{J}}^{\rm nPhys}_Q$, which would be reflected in $\Delta J_{Q}$.

Therefore, it is expected that the heat flux regularization leads to converse behavior in the RMSEs and $\Delta J_{Q}$.
In addition, since the regularization plays a role in suppressing the overfitting and underfitting of the atomic components, 
it improves the robustness of the ANN potentials.
The above discussion also suggests that it is insufficient to measure the robustness of ANN potentials by the RMSEs alone.
Minimizing the RMSEs does not necessarily mean constructing ANN potentials with high robustness because either overfitting or underfitting of the atomic components may have occurred.
$\Delta J_{Q}$ should be used as an indicator of the robustness, in addition to the RMSEs.

\subsection{\label{ssec:FPMDresults} 
Results of ANN potentials trained with FPMD data}
The investigation using the EIP of Ag$_2$Se discussed in the previous section~\ref{ssec:EIPresults} indicated the procedure for estimating an appropriate $p_J$ without reference data using the variations of the RMSEs and $\Delta J_Q$ with respect to $p_J$.
This section applies the same approach to Ag$_2$S.

\subsubsection{\label{sssec:re-baAg2S}Power spectra of $\beta$- and $\alpha$-Ag$_2$S and the selection of appropriate $p_J$} 
\begin{figure*}[b]
\begin{center}
  \includegraphics[width=15cm]{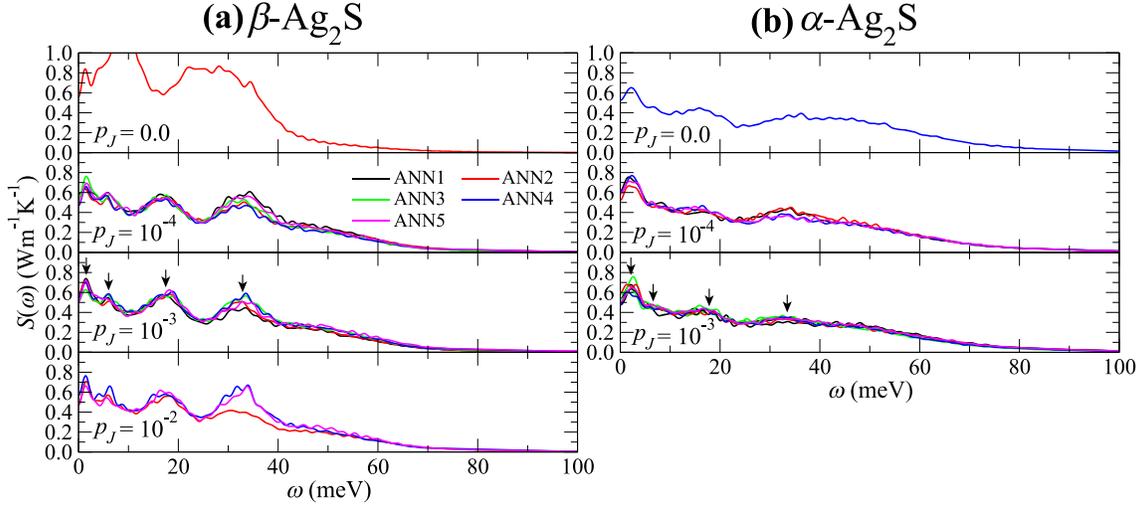}
  \caption{\label{Fig5} $S(\omega)$ obtained by the trained ANN potentials with heat flux regularization under each $p_J$ for (a) $\beta$- and (b) $\alpha$-Ag$_2$S. ANN potentials (ANN1-ANN5) are trained by five different initial weight parameters (identified by colors in the figures above). The black arrows in the power spectra for $p_J$ = 10$^{\mathchar`-3}$ indicate the peak positions of the $\beta$ phase, which are discussed in section~\ref{sssec:dif_spectra}.}
\end{center}
\end{figure*}

The power spectra for the ANN potentials for $\beta$- and $\alpha$-Ag$_2$S are shown in Figs.~\ref{Fig5}(a) and \ref{Fig5}(b), respectively.
The corresponding HFACFs and $\kappa(t)$ are shown in Fig.~S6 of the SM.
With $p_J$ = 0.0, only one ANN potential can perform stable $NVE$-MD simulations for 1,000,000 steps for each of $\beta$- and $\alpha$-Ag$_2$S.
However, with $p_J$ = 10$^{\mathchar`-4}$ or 10$^{\mathchar`-3}$, stable simulations could be performed with almost all five ANN potentials.
Only one ANN potential provided a failed simulation: ANN3 with $p_J = 10^{\mathchar`-4}$ for $\alpha$ phase was failed.
The different ANN potentials displayed very similar spectral profiles.
These results show that heat flux regularization improved the robustness of the ANN potentials.
However, with a larger $p_J$ = 10$^{\mathchar`-2}$, only three (ANN2, ANN4, and ANN5) and zero out of five ANN potentials provided stable MD simulations for $\beta$- and $\alpha$-Ag$_2$S, respectively.

\begin{figure}[h]
\begin{center}
  \includegraphics[width=7.0cm]{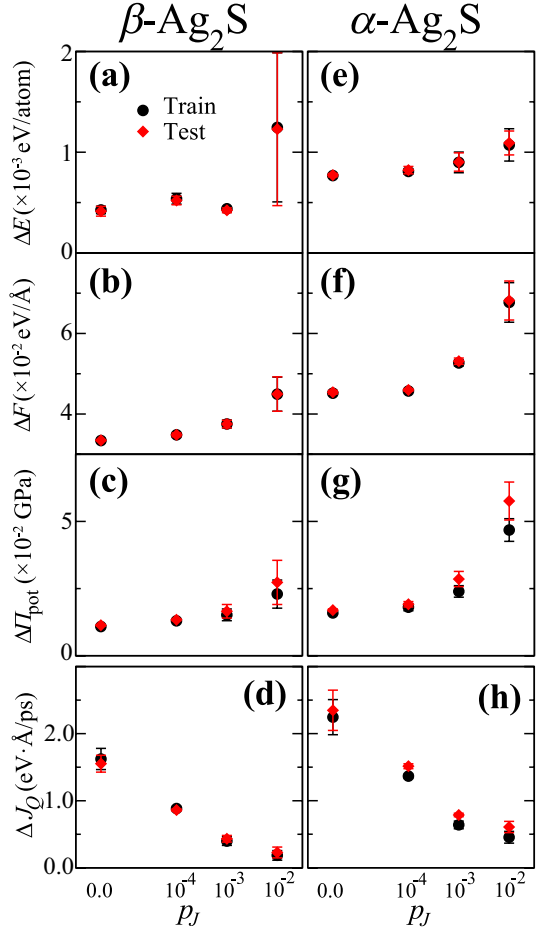}
  \caption{\label{Fig6} Averaged RMSEs and corresponding standard deviations (error bars) of $\Delta E$, $\Delta F$, and $\Delta \varPi_{\rm pot}$ over the five trained ANN potentials for (a-c) $\beta$- and (e-g) $\alpha$-Ag$_2$S.
  The values of $\Delta J_Q$ for (d) $\beta$- and (h) $\alpha$-Ag$_2$S are also shown.
  Black circles and red diamonds represent results obtained for the Train and Test data sets, respectively.}
\end{center}
\end{figure}

The RMSEs (i.e., $\Delta E$, $\Delta F$, and $\Delta \varPi_{\rm pot}$) of $\beta$- and $\alpha$-Ag$_2$S are shown in Figs.~\ref{Fig6}(a)-\ref{Fig6}(c) and~\ref{Fig6}(e)-\ref{Fig6}(g), respectively.
In addition, the values of the magnitude of the heat flux for the regularization $\Delta J_Q$ for $\beta$ and $\alpha$ phases are shown in Figs.~\ref{Fig6}(d) and~\ref{Fig6}(h), respectively.
In both phases, the averages and error bars of the RMSEs increased considerably as $p_J$ increased from 10$^{\mathchar`-3}$ to 10$^{\mathchar`-2}$.
The values of $\Delta J_Q$ for $p_J$ = 10$^{\mathchar`-3}$ were substantially smaller than those for $p_J < 10^{\mathchar`-3}$.
$p_J$ = 10$^{\mathchar`-3}$ was therefore considered to be the most appropriate value for both phases.
It was confirmed that the $g$($r$) and the MSDs obtained by the FPMD simulation and the ANN potentials trained with $p_J$ = 10$^{\mathchar`-3}$ coincided fairly well, as shown in Figs.~S7 and~S8 of the SM.

We hypothesize that the one reason of the difficulty of successfully performing
long MD simulations of ANN potentials with $p_J$ = 0.0 for both $\beta$- and $\alpha$-Ag$_2$S is because of the large value of $\Delta J_Q$.
The magnitudes with $p_J$ = 0.0 for both phases were much larger than those for the regularized trainings (i.e., $p_J$ > 0.0), suggesting that the ANN potentials provided larger values of ${\bf{J}}^{\rm nPhys}_Q$ as well.
Therefore, the ANN potentials trained without regularization are considered to have undergone significant overfitting or underfitting of the atomic components (as explained in section~\ref{sssec:val_heat}), making it difficult to predict accurate physical quantities; in particular, the atomic forces, thereby causing fatal computing errors to occur locally during MD simulations.
Note that another reason could be attributed to the low diversity of training data.
Because the FPMD training data were generated by a shorter simulation than the EIP-MD data, the diversity of atomic configurations included should be much lower.
It was difficult to construct a robust ANN potential that could perform long-term MD simulations using such a low diversity data.~\cite{Shimamura_2019}

\subsubsection{\label{sssec:dif_spectra}
Power spectrum analysis for $\beta$ and $\alpha$ phases}
We compared power spectra of $\beta$- and $\alpha$-Ag$_2$S for the most appropriate value $p_J = 10^{\mathchar`-3}$, as shown in Fig.~\ref{Fig5}.
The TCs of $\beta$- and $\alpha$-Ag$_2$S were 0.541 $\pm$ 0.030 and 0.516 $\pm$ 0.050 Wm$^{\mathchar`-1}$K$^{\mathchar`-1}$, respectively.
These values are in close agreement with each other and with the experimental values.~\cite{Hirata_2020,Matsunaga_2021}

\begin{figure*}[h]
\begin{center}
  \includegraphics[width=16cm]{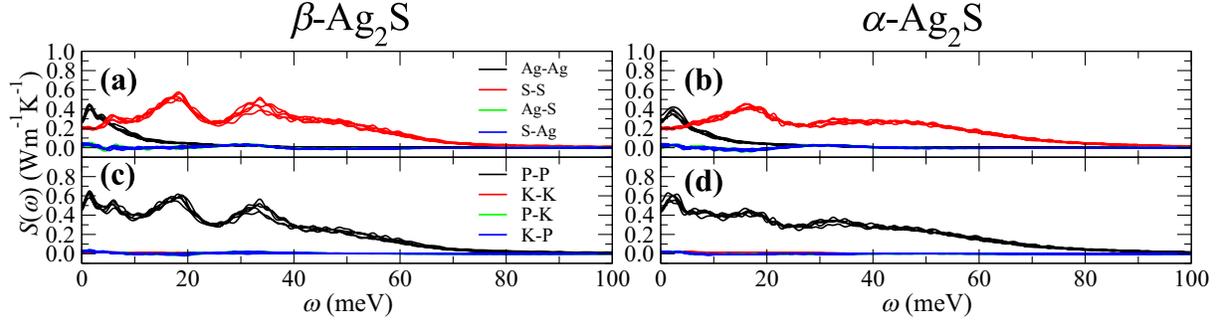}
  \caption{\label{Fig7} Decomposed $S(\omega)$ by (a-b) elemental contributions and (c-d) kinetic/potential parts obtained by the ANN potentials trained with the appropriate regularization for $\beta$- and $\alpha$-Ag$_2$S.
  The former is expressly $S$ = $S_{\rm Ag\mathchar`-Ag}$ + $S_{\rm Se\mathchar`-Se}$ + $S_{\rm Ag\mathchar`-Se}$ + $S_{\rm Se\mathchar`-Ag}$ and the latter $S$ = $S_{\rm P\mathchar`-P}$ + $S_{\rm K\mathchar`-K}$ + $S_{\rm P\mathchar`-K}$ + $S_{\rm K\mathchar`-P}$. For both $\beta$- and $\alpha$-Ag$_2$S, $p_J$ = 10$^{\mathchar`-3}$ is used for the regularization. For each decomposed $S(\omega)$, results obtained from the five ANN potentials trained with different initial weight parameters are shown.}
\end{center}
\end{figure*}

However, the profiles of $S(\omega)$ show different features.
The peaks at 6 and 34 meV present in the $\beta$-Ag$_2$S spectrum were no longer visible in the spectrum of $\alpha$-Ag$_2$S, and the height of the peak at 18 meV is lower in the latter spectrum.
In contrast, $\alpha$-Ag$_2$S was found to have a heavier tailed distribution up to 90 meV.
The components of the power spectra defined in section~\ref{sssec:ps} are displayed in Fig.~\ref{Fig7}.
From the power spectra of the elemental contribution for $\beta$- and $\alpha$-Ag$_2$S shown in Figs.~\ref{Fig7}(a) and~\ref{Fig7}(b), respectively, the peaks at 6, 18, and 34 meV and the long tail of the spectra up to 90 meV are mainly attributed to $S_{\rm S\mathchar`-S}(\omega)$.
Conversely, the profiles of $S_{\rm Ag\mathchar`-Ag}(\omega)$ and the cross terms ($S_{\rm Ag\mathchar`-S}(\omega)$ and $S_{\rm S\mathchar`-Ag}(\omega)$) did not display substantial variation between the $\beta$ and $\alpha$ phases.
The peak at 3 meV resulted from $S_{\rm Ag\mathchar`-Ag}(\omega)$ for both phases.
From the power spectra of the kinetic/potential parts shown in Figs.~\ref{Fig7}(c) and ~\ref{Fig7}(d), the potential part ($S_{\rm P\mathchar`-P}(\omega)$) dominates the spectrum in both phases.
The power spectrum describes the frequency distribution of phonons, and the intensity of its peak corresponds to the contribution of individual phonons.~\cite{Mahdizadeh_2014,Zhang_2013}
In addition, the phonon frequency analysis results for $\beta$-Ag$_2$S at 300 K using the Boltzmann transport equation~\cite{Zhou_2020} suggest that acoustic phonons predominantly contribute to the power spectrum up to $\sim$5 meV, while optical phonons predominantly contribute to the spectrum above that frequency.
Considering these findings, our results for $\alpha$-Ag$_2$S suggest that instead of a decrease in optical phonons corresponding to the peaks at 6, 18, and 34 meV, high-frequency optical phonons with energies up to 90 meV appeared.
The frequency ranges of the optical phonons due to the interactions between S atoms were different for $\beta$ and $\alpha$ phases.
The heat flux regularization approach will allow us to access information about such phonon distributions that was previously unavailable in MLIPs.

\section{\label{sec:conclusion}Conclusion}
In this paper, we proposed a data-driven approach to reduce the nonphysical heat flux (i.e., ${\bf{J}}^{\rm nPhys}_Q$) caused by the $atomic$ $gauge$~\cite{Kim_2021} by incorporating the regularization of the heat flux into the training of MLIPs.
Because adjusting the strength of regularization is the key to achieving this reduction, we investigated the conditions for constructing MLIPs under which the power spectra of the heat flux of $\beta$- and $\alpha$-Ag$_2$Se calculated by the EIP of Ag$_2$Se could be reproduced.
The EIP was composed of only pairwise functions where ${\bf{J}}^{\rm nPhys}_Q$ did not occur.
The ANN potential was adopted as the MLIP in this study.

The strength of the effects could be controlled by the coefficient $p_J$.
We found that the optimal value of $p_J$ could be estimated without access to reference spectra by clarifying the correlation of the RMSEs (i.e., $\Delta E$, $\Delta F$, and $\Delta \varPi_{\rm pot}$) and the magnitude of the heat flux for regularization (i.e., $\Delta J_Q$) with respect to $p_J$; that value of $p_J$ that minimizes both $\Delta J_{Q}$ and the RMSEs should be selected.

The heat flux regularization was applied to the ANN potentials trained with the FPMD data of $\beta$- and $\alpha$-Ag$_2$S.
From their power spectra, the frequency ranges of optical phonons associated with due to the interactions between S atoms was found to be different in the two phases.

We also found that reducing ${\bf{J}}^{\rm nPhys}_Q$ through the regularization improved the robustness of the ANN potentials.
This occurred because the regularization suppressed the overfitting and underfitting of the atomic components (i.e., $\varepsilon^{\rm ANN}_i$, ${\bf F}^{\rm ANN}_{ij}$, and W$^{\rm ANN}_i$).
We identified $\Delta J_{Q}$ as an useful indicator for measuring the robustness, in addition to the RMSEs.

With ANN potentials trained with the heat flux regularization, in future, we plan to investigate the composition dependence of the frequency distribution of phonons and the thermal conductivity of silver chalcogenide mixtures such as Ag$_2$S$_{1\mathchar`-x}$Se$_x$~\cite{Matsunaga_2021}, using power spectrum analysis and effective analytical techniques~\cite{Fujii_2020,Fujii_2018,Zhou_2015,Zhou_2015_2,Saaskilahti_2014,Ouyang_2014} based on the decomposition of heat flux.

\section*{Supplementary Material}
See Supplemental Materials for detailed descriptions regarding the EIP of Ag$_2$Se (Section I), training of ANN potentials (Section II), and the definition of cumulative TC, $\kappa(t)$ (Section III).
In Fig.~S1, we show the decomposed power spectra according to their elemental contributions and kinetic/potential parts obtained by the EIP of Ag$_2$Se and ANN potentials trained with the appropriate regularization, for $\beta$- and $\alpha$-Ag$_2$Se.
In Figs.~S2,~S3,~S4, and~S5, we show the HFACFs, $\kappa(t)$, $g(r)$, and MSDs obtained by the EIP of Ag$_2$Se and trained ANN potentials with the heat flux regularization under each $p_J$ for $\beta$- and $\alpha$-Ag$_2$Se.
In Fig.~S6, we show the HFACFs and $\kappa(t)$ obtained by trained ANN potentials with the regularization under each $p_J$, for $\beta$- and $\alpha$-Ag$_2$S.
In addition, the $g(r)$ and the MSDs obtained by the FPMD simulation and the trained ANN potentials are shown in Figs.~S7 and~S8.

\begin{acknowledgments}
We thank Shogo Fukushima and Yusuke Takeshita for providing the codes for calculating thermal conductivity and for teaching the ANN potential training technique.
This study was supported by MEXT/JSPS KAKENHI Grant Numbers Nos. 21H01766 and 19K14676, and JST CREST Grant Number JPMJCR18I2, Japan. The authors thank the Supercomputer Center, the Institute for Solid State Physics, University of Tokyo for the use of the facilities. 
The computations were also carried out using the facilities of the Research Institute for Information Technology, Kyushu University.
\end{acknowledgments}

\section*{Data availavility}
The data that support the findings of this study are available from the corresponding author upon reasonable request.

\clearpage
\section*{Figures}

%\end{document}

\clearpage
%\documentclass[preprint,review,12pt]{revtex4}
%\usepackage{amsmath}
%\usepackage{color}
%\usepackage{graphicx}
%\usepackage{multirow}
%\usepackage{array}
%\renewcommand{\figurename}{Figure S}
%\renewcommand{\tablename}{TABLE S}
%\newcolumntype{C}[1]{>{\hfil}m{#1}<{\hfil}}
%\makeatletter
%　　\renewcommand{\theequation}{
%    \thesection.\arabic{equation}}
%    \@addtoreset{equation}{section}
%\makeatother

%\begin{document}
%\title{Supplementary Material for ``Construction of Machine-Learning Interatomic Potential Under Heat Flux Regularization and Its Application to Power Spectrum Analysis for Silver Chalcogenides''}

%\author{Kohei Shimamura}
%\email{shimamura@kumamoto-u.ac.jp.}
%\affiliation{Department of Physics, Kumamoto University, Kumamoto 860-8555, Japan}

%\author{Akihide Koura}
%\affiliation{Department of Physics, Kumamoto University, Kumamoto 860-8555, Japan}

%\author{Fuyuki Shimojo}
%\affiliation{Department of Physics, Kumamoto University, Kumamoto 860-8555, Japan}

%\maketitle

%\setlength{\baselineskip}{24pt}

{\Large{Supplementary Material for ``Construction of Machine-Learning Interatomic Potential Under Heat Flux Regularization and Its Application to Power Spectrum Analysis for Silver Chalcogenides''}}
\setcounter{section}{0}
\setcounter{figure}{0}
\setcounter{equation}{0}
\renewcommand{\figurename}{Figure S}
\renewcommand{\tablename}{TABLE S}

\clearpage
\section{Empirical Interatomic Potential of Ag$_2$Se and definition of its atomic potential, force, and virial tensor}
The empirical interatomic potential $V^{\rm EIP}(r)$ for Ag$_2$Se proposed by Rino {\it{et al.}}~\cite{Rino_1988_S} is composed of three pairwise functions: repulsive, dielectric dipole interaction, and Coulomb interaction terms
\begin{eqnarray}
V^{\rm EIP}\left( r \right) = &
  \begin{cases}
    \frac{0.2408}{r^{11}} + \frac{0.2025}{r} &\left( {\rm Ag\mathchar`-Ag} \right)\\
    \frac{86.6614}{r^9} -\frac{0.7088{\rm exp}(-r/4.43)}{r^4} - \frac{0.405}{r} &\left( {\rm Ag\mathchar`-Se} \right)\\
    \frac{220.1905}{r^7} -\frac{5.67{\rm exp}(-r/4.43)}{r^4} + \frac{0.81}{r} &\left( {\rm Se\mathchar`-Se} \right)
  \end{cases}
\label{eq:fc},
\end{eqnarray}
where the interatomic distance $r$ is measured in {\AA} and the energies are given in units of ${e^2/}${\AA} = 14.389 eV.
In addition, the functional form is defined for each component pair, i.e., Ag-Ag, Ag-Se, and Se-Se.
For the empirical interatomic potential, atomic potential energy of the $i$th atom $\varepsilon^{\rm EIP}_i$ can be written as
\begin{eqnarray}
\varepsilon^{\rm EIP}_i = \frac{1}{2}\sum_{j \ne i}V^{\rm EIP}(r_{ij})
\label{eq:ape},
\end{eqnarray}
where $r_{ij}$ $=$ $|{\bf{r}}_{ij}|$ $=$ $|{\bf{r}}_{j} - {\bf{r}}_{i} |$ denotes the interatomic distance between the $i$th and $j$th atoms.
The summation $\sum^{N_{\rm atom}}_i \varepsilon^{\rm EIP}_i$ is equal to total potential energy $E^{\rm EIP}$.
$N_{\rm atom}$ denotes the number of atoms in the system.
Atomic virial tensor for the empirical interatomic potential ${\rm W}^{\rm EIP}_i$ is given by
\begin{eqnarray}
{\rm W}^{\rm EIP}_i = -\frac{1}{2}\sum_{j \ne i} {\bf{r}}_{ij} \otimes {\bf{F}}^{\rm EIP}_{ij}
\label{eq:avt}, 
\end{eqnarray}
with ${\bf{F}}^{\rm EIP}_{ij} = \frac{\partial V^{\rm EIP}(r_{ij})}{\partial{\bf{r}}_{ij}}$, and the summation $\sum^{N_{\rm atom}}_j {\bf{F}}^{\rm EIP}_{ij}$ is equal to atomic force ${\bf{F}}^{\rm EIP}_i$.
The sum of ${\rm W}^{\rm EIP}_i$ corresponds to the total virial tensor W$^{\rm EIP}$.
\begin{eqnarray}
{\rm W}^{\rm EIP} = \sum^{N_{\rm atom}}_{i} {\rm W}^{\rm EIP}_i
\label{eq:sumv}.
\end{eqnarray}

\clearpage
\section{Training of ANN Potentials in this study}
\subsection{Training with Data Generated from Empirical Interatomic Potential of Ag$_2$Se}
The total number of training data used for the cost function (Eq.~(8) in the main text) reached $N_I \times (D_{\rm Energy} + D_{\rm Force}+D_{\rm Virial}+D_{\rm Velocity}) = 1000 \times ( 1 + 3 \times 384 + 6 + 3 \times 384) = 2311000$, where $N_I$ is the number of MD steps, and $D_{\rm Energy}$, $D_{\rm Force}$, $D_{\rm Virial}$, and $D_{\rm Velocity}$ are the dimensions of potential energy, atomic force, virial, and atomic velocity in one MD step data, respectively.
Note that how to generate these data was explained in section II C in the main text.

The number of epochs was set to 50 for all training with data generated by the empirical interatomic potential (EIP) of Ag$_2$Se, and $p_E = p_F = 1.0$ and $p_W = 10^{\mathchar`-5}$ during the training.
Please see section III B in the main text for the values of $p_J$ in each training.

\subsection{Training with FPMD Data}
The total number of training data used for the cost function (Eq.~(8) in the main text) reached $N_I \times (D_{\rm Energy} + D_{\rm Force}+D_{\rm Virial}+D_{\rm Velocity}) = 1000 \times ( 1 + 3 \times 384 + 6 + 3 \times 384) = 2311000$.
Note that how to generate these data was explained in section II D in the main text.

For all training with FPMD data, we took two-step scheme to train ANN potential efficiently by manipulating $p_E$, $p_F$, and $p_W$.
$p_F$ was set to $1.0$ in both steps, while $p_E = 10^{\mathchar`-3}$ and $p_W = 10^{\mathchar`-6}$ in the first step and $p_E = 10^{\mathchar`-1}$ and $p_W = 10^{\mathchar`-1}$ in the second step.
The number of epochs was set to 20 for each step.
Adjusting the coefficients of the cost function during training in this way for efficiency has been actively carried out in previous studies~\cite{Takeshita_2022_S,Shimamura_2021_S,Lee_2020_S,Shimamura_2019_S,Zhang_2018_S}.
In addition, the two-step scheme adopted in this study above has been found to be capable of constructing highly accurate ANN potentials trained with Ag$_2$Se FPMD data~\cite{Takeshita_2022_S}.
Please see section III C in the main text for the values of $p_J$ in each training.

\subsection{The formulas of root mean square errors}
\noindent
For total potential energy,
\begin{eqnarray}
\Delta E &=& \sqrt{\frac{1}{N_I}\sum^{N_I}_{I} \left(  \frac{   E^{\rm ANN}_I - E^{\rm Ref}_I  }{N_{{\rm atom},I}} \right)^2}. \nonumber
\end{eqnarray}
For atomic force,
\begin{eqnarray}
\Delta F &=& \sqrt{\frac{1}{N_I}\sum^{N_I}_{I} \frac{1}{3N_{{\rm atom},I}} \sum^{N_{{\rm atom},I}}_{i}\ 
    \left({\bf{F}}^{\rm ANN}_{I,i} - {\bf{F}}^{\rm Ref}_{I,i} \right)^2}. \nonumber
\end{eqnarray}
For virial stress from the contribution of the potential energy,
\begin{eqnarray}
\Delta \varPi_{\rm pot} &=& \sqrt{\frac{1}{N_I}\sum^{N_I}_{I} \frac{1}{6\varOmega^{\rm Ref}_I} \sum^{6}_{j}\ 
    \left(  {W}^{\rm ANN}_{I,j} - {W}^{\rm Ref}_{I,j} \right)^2}. \nonumber
\end{eqnarray}

\clearpage
\section{Cumulative Thermal Conductivity}
The cumulative thermal conductivity $\kappa(t)$ is calculated by
\begin{eqnarray}
\kappa(t) &=& \frac{1}{3k_{\rm B}T^2\varOmega}\int^{t}_{0} \langle{\bf{J}}_Q(t') \cdot {\bf{J}}_Q(0)\rangle dt'
\label{eq:ckappa},
\end{eqnarray}
where $k_{\rm B}$, $T$, $\varOmega$, and ${\bf{J}}_Q$ are Boltzmann constant, temperature, volume of supercell, and heat flux in Eq.~(5) of the main text, respectively.
In this study, $\kappa(t)$ were computed through Eq.~(\ref{eq:ckappa}) with the upper limit of integration of 2 ps.
This is because $\kappa(t)$ converged well at 2 ps, as shown in Figs.~S\ref{FigS2},~S\ref{FigS3}, and~S\ref{FigS6}.

\clearpage
\section{Supplementary Figures}
\begin{figure}[h]
\begin{center}
  \includegraphics[width=16cm]{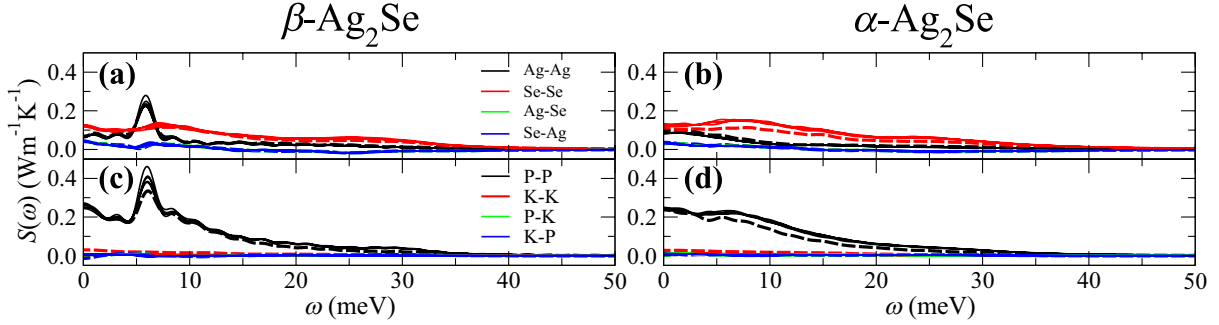}
  \caption{\label{FigS1} Decomposed power spectra of heat flux $S(\omega)$ by (a-b) elemental contributions and (c-d) kinetic/potential parts obtained by the empirical interatomic potential (EIP) of Ag$_2$Se (dashed curves) and ANN potentials trained with regularization (solid curves) for $\beta$- and $\alpha$-Ag$_2$Se.
  The former is expressly $S$ = $S_{\rm Ag\mathchar`-Ag}$ + $S_{\rm Se\mathchar`-Se}$ + $S_{\rm Ag\mathchar`-Se}$ + $S_{\rm Se\mathchar`-Ag}$ and the latter $S$ = $S_{\rm P\mathchar`-P}$ + $S_{\rm K\mathchar`-K}$ + $S_{\rm P\mathchar`-K}$ + $S_{\rm K\mathchar`-P}$. For $\beta$- and $\alpha$-Ag$_2$Se, $p_J$ = 5$\times$10$^{\mathchar`-3}$ and 10$^{\mathchar`-3}$ are used for regularization, respectively. For each decomposed $S(\omega)$, results obtained from the five ANN potentials trained with different initial weight parameters are shown.}
\end{center}
\end{figure}

\begin{figure}[h]
\begin{center}
  \includegraphics[width=16cm]{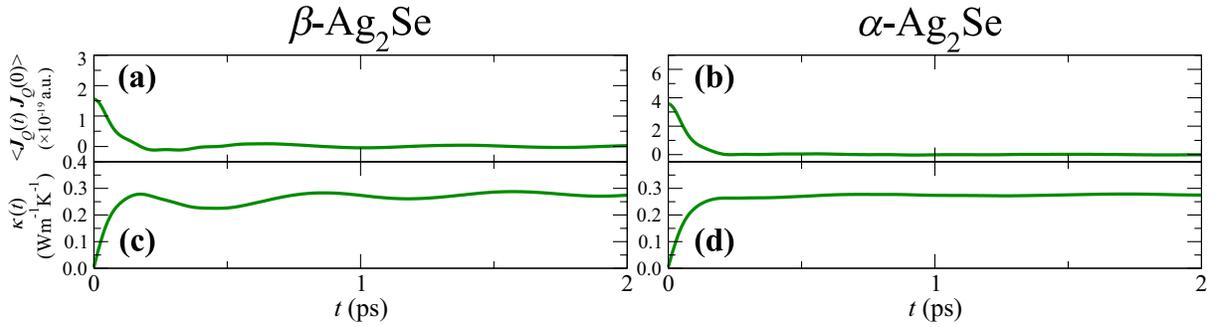}
  \caption{\label{FigS2} (a-b) Autocorrelation functions of heat flux $\langle{\bf{J}}_Q(t) \cdot {\bf{J}}_Q(0)\rangle$ (atomic unit) and (c-d) cumulative thermal conductivities $\kappa(t)$ defined in Eq.~(\ref{eq:ckappa}) (Wm$^{-1}$K$^{-1}$) up to 2 ps obtained by the EIP of Ag$_2$Se  for $\beta$- and $\alpha$-Ag$_2$Se.}
\end{center}
\end{figure}

\begin{figure}[th]
\begin{center}
  \vspace{-0.5cm}
  \includegraphics[width=14cm]{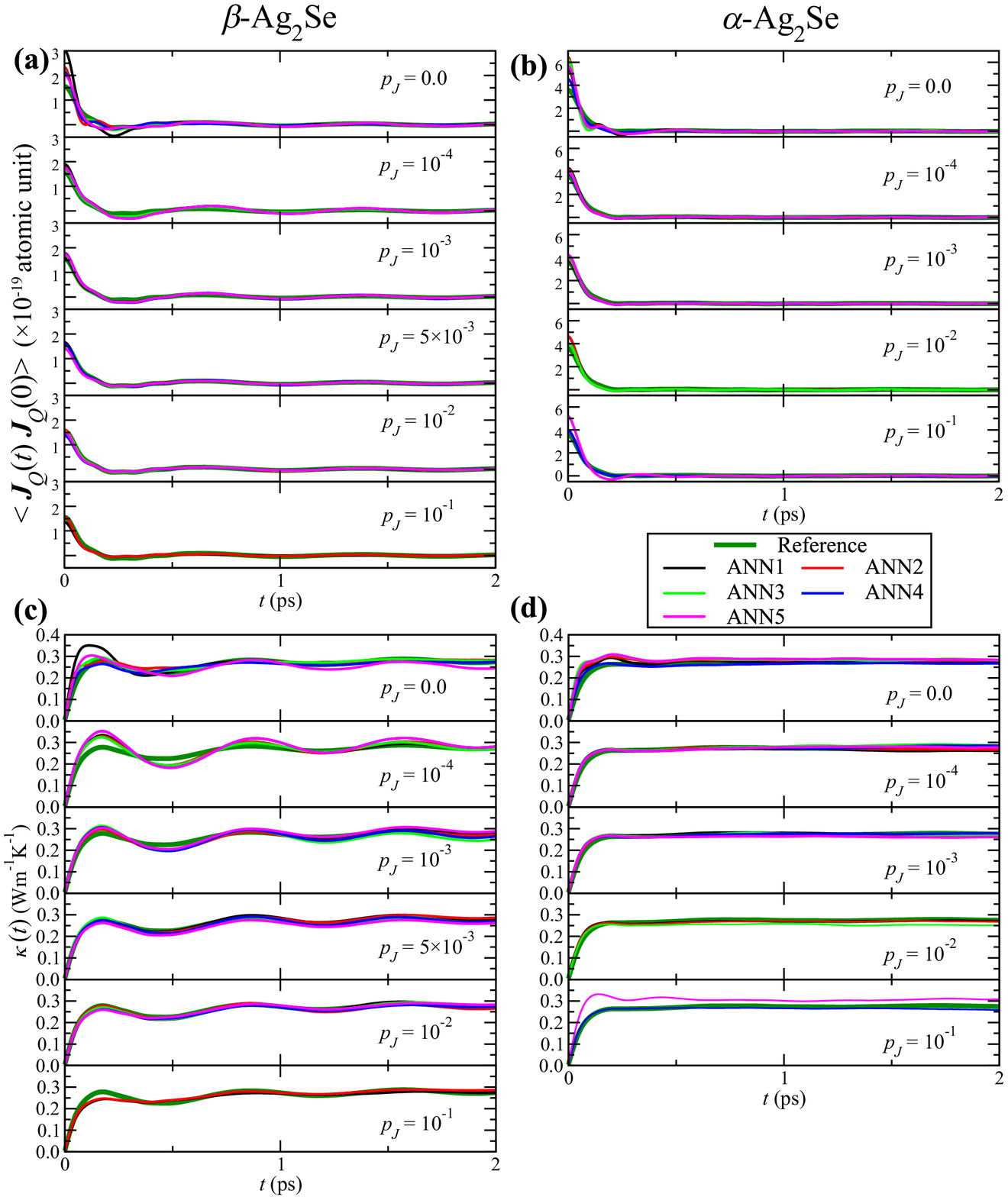}
  \caption{\label{FigS3} (a-b)  $\langle{\bf{J}}_Q(t) \cdot {\bf{J}}_Q(0)\rangle$ (atomic unit) and (c-d) $\kappa(t)$ (Wm$^{\mathchar`-1}$K$^{\mathchar`-1}$) obtained by the EIP of Ag$_2$Se (Reference, dark green) and trained ANN potentials with heat flux regularization under each $p_J$ for $\beta$- and $\alpha$-Ag$_2$Se. ANN potentials (ANN1-ANN5) are trained with five different initial weight parameters (identified by colors in the figures above).}
\end{center}
\end{figure}

\begin{figure}[h]
\begin{center}
  \includegraphics[width=15cm]{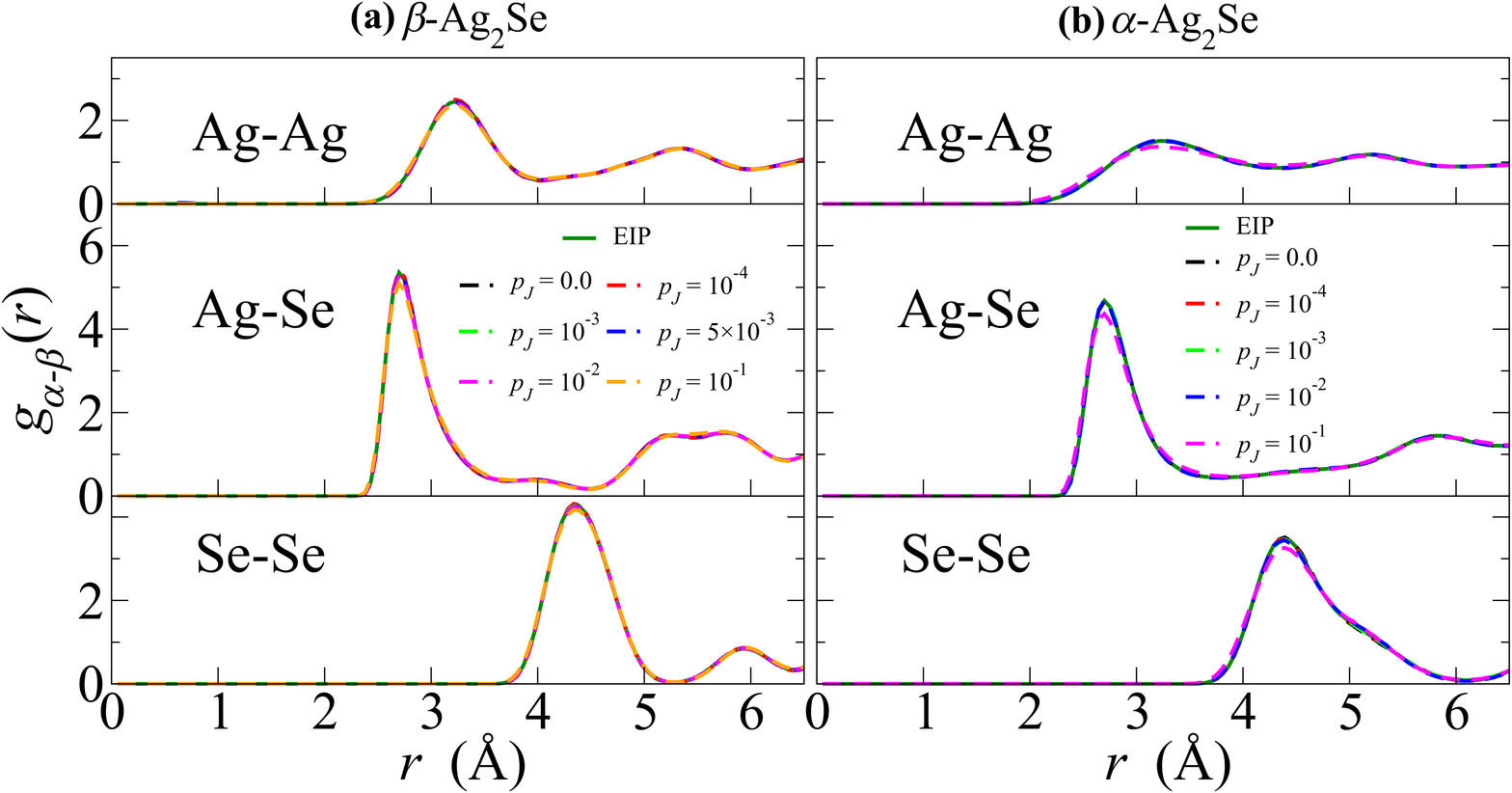}
  \caption{\label{FigS4} Partial radial distribution functions $g_{\alpha \mathchar`-\beta}$($r$) for atomic pairings Ag-Ag, Ag-Se, and Se-Se calculated by MD simulations with the EIP of Ag$_2$Se and ANN potentials with the regularization of each $p_J$ for (a) $\beta$- and (b) $\alpha$-Ag$_2$Se. For each $p_J$, the one giving the worst training accuracy for the atomic force was used from the ANN potentials for which the power spectrum could be calculated.}
\end{center}
\end{figure}

\begin{figure}[h]
\begin{center}
  \includegraphics[width=15cm]{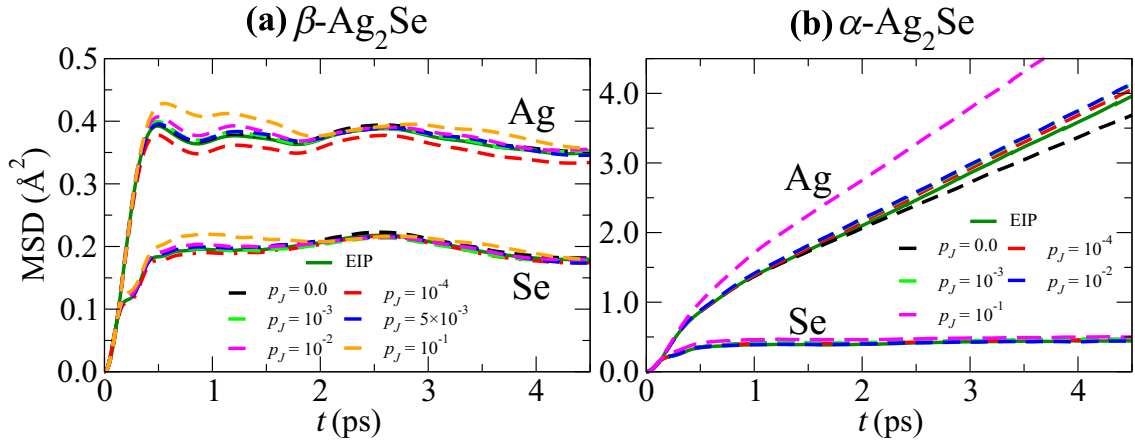}
  \caption{\label{FigS5} Mean squared displacements (MSDs) (\AA$^2$) of Ag and Se atoms calculated by MD simulations using the EIP of Ag$_2$Se and ANN potentials with the regularization of each $p_J$ for (a) $\beta$- and (b) $\alpha$-Ag$_2$Se. For each $p_J$, the one giving the worst training accuracy for the atomic force was used from the ANN potentials for which the power spectrum could be calculated.}
\end{center}
\end{figure}

\begin{figure}[h]
\begin{center}
  \includegraphics[width=15cm]{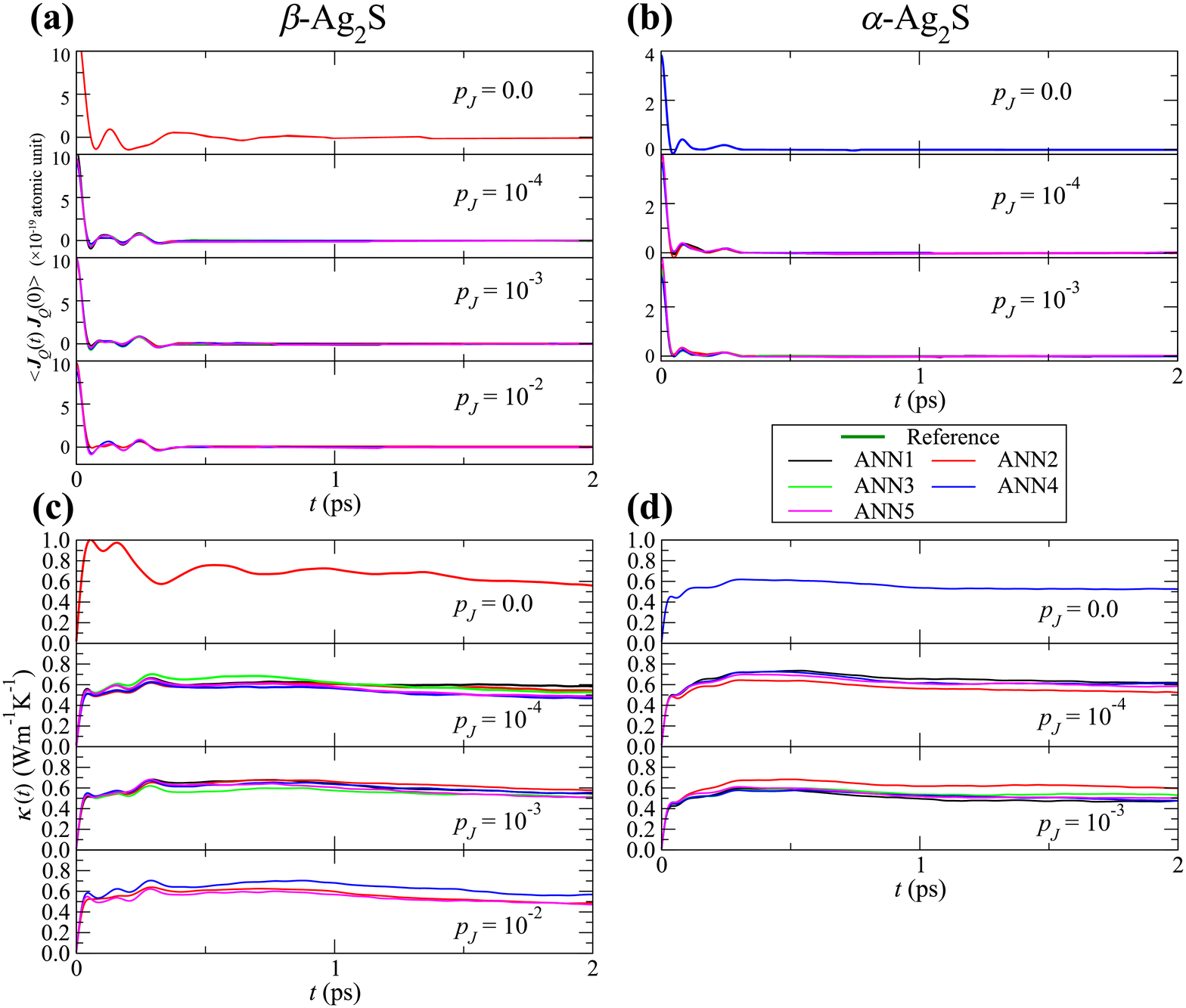}
  \caption{\label{FigS6} (a-b) $\langle{\bf{J}}_Q(t) \cdot {\bf{J}}_Q(0)\rangle$ (atomic unit) and (c-d) $\kappa(t)$ (Wm$^{\mathchar`-1}$K$^{\mathchar`-1}$) obtained by the trained ANN potentials with heat flux regularization under each $p_J$ for $\beta$- and $\alpha$-Ag$_2$S. ANN potentials (ANN1-ANN5) are trained with five different initial weight parameters (identified by colors in the figures above).}
\end{center}
\end{figure}

\begin{figure}[h]
\begin{center}
  \includegraphics[width=15cm]{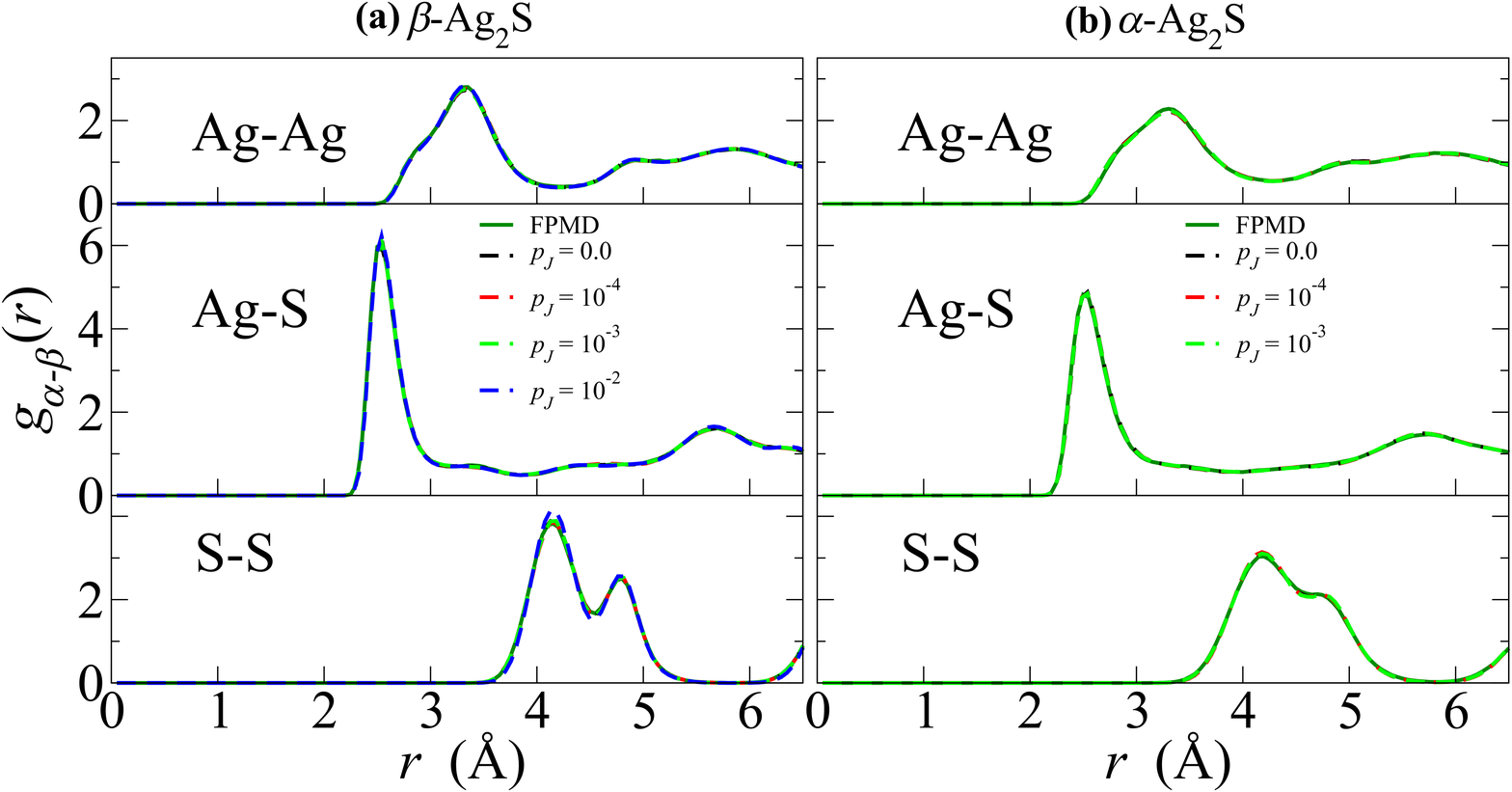}
  \caption{\label{FigS7} $g_{\alpha \mathchar`-\beta}$($r$) for atomic pairings Ag-Ag, Ag-S, and S-S calculated in the FPMD simulation and MD simulations with ANN potentials with the regularization of each $p_J$ for (a) $\beta$- and (b) $\alpha$-Ag$_2$S. For each $p_J$, the one giving the worst training accuracy for the atomic force was used from the ANN potentials for which the power spectrum could be calculated.}
\end{center}
\end{figure}

\begin{figure}[h]
\begin{center}
  \includegraphics[width=15cm]{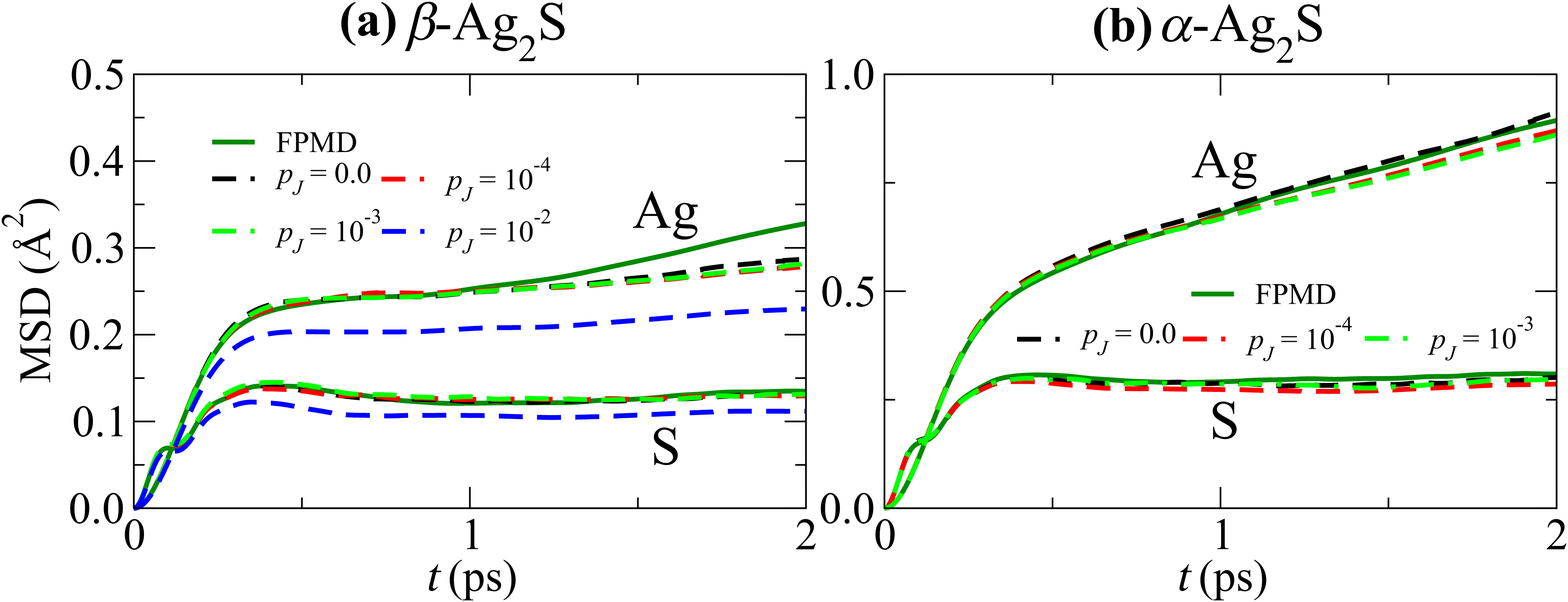}
  \caption{\label{FigS8} MSDs (\AA$^2$) of Ag and S atoms calculated in the FPMD simulation and MD simulations using the ANN potentials with the regularization of each $p_J$ for (a) $\beta$- and (b) $\alpha$-Ag$_2$S. For each $p_J$, the one giving the worst training accuracy for the atomic force was used from the ANN potentials for which the power spectrum could be calculated.}
\end{center}
\end{figure}

%\bibliography{Supplementary.bib}% Produces the bibliography via BibTeX.

%\end{document}

\end{document}